\documentclass[5p,times]{elsarticle}
\def\tsc#1{\csdef{#1}{\textsc{\lowercase{#1}}\xspace}}
\tsc{WGM}
\tsc{QE}

\usepackage{amsmath,amssymb,amsfonts}
\usepackage{graphicx}
\usepackage{textcomp}
\usepackage{xcolor}
\usepackage{multirow}
\usepackage{float}
\usepackage{booktabs}
\usepackage{subcaption}
\usepackage{multirow}
\usepackage{threeparttable}
\usepackage{tabularx}

\usepackage{mathtools}

\usepackage[ruled,vlined,linesnumbered]{algorithm2e}
\SetCommentSty{scriptsize}
\SetKwComment{tcc}{[}{]}  %
\SetKwIF{If}{ElseIf}{Else}{if}{}{else if}{else}{endif}
\SetKwFor{While}{while}{}{endw}
\SetKwFor{ForEach}{for each}{}{endfch}
\SetKwFor{For}{for}{}{}%
\SetKwComment{tcp}{$\rightarrow$\quad}{}
\usepackage{etoolbox}
\usepackage{xstring}
\usepackage{calc}

\newlength{\xalgowidth}
\newlength{\xalgoremainder}
\newlength{\xindentwidth}

\newenvironment{vAlgorithm*}[3][]{%
  \setlength{\xalgowidth}{#2} %
  \setlength{\xindentwidth}{#3} %
  \setlength{\xalgoremainder}{\textwidth-\xalgowidth} %
  \SetCustomAlgoRuledWidth{\xalgowidth} %
  \IncMargin{\xindentwidth}
  \begin{algorithm*}[#1]
}%
{%
  \end{algorithm*} 
  \DecMargin{\xindentwidth}
}%

\newenvironment{vAlgorithm}[3][]{%
  \setlength{\xalgowidth}{#2} %
  \setlength{\xindentwidth}{#3} %
  \setlength{\xalgoremainder}{\columnwidth-\xalgowidth} %
  \SetCustomAlgoRuledWidth{\xalgowidth} %
  \IncMargin{\xindentwidth}
  \begin{algorithm}[#1] %
}%
{%
  \end{algorithm} %
  \DecMargin{\xindentwidth}
}%

\makeatletter
\patchcmd{\@algocf@start}{%
\begin{lrbox}{\algocf@algobox}%
}{%
\rule{0.5\xalgoremainder}{\z@}%
\begin{lrbox}{\algocf@algobox}%
\begin{minipage}{\xalgowidth}%
}{}{}
\patchcmd{\@algocf@finish}{%
\end{lrbox}%
}{%
\end{minipage}%
\end{lrbox}%
}{}{}
\makeatother

\newlength{\commentWidth}
\usepackage{enumitem}
\usepackage{array}
\usepackage{setspace}

\PassOptionsToPackage{hyphens}{url} %
\usepackage{hyperref}
\hypersetup{%
    unicode=true,       %
    pdftitle={Characterizing the Performance of Node-Aware Strategies for Irregular Point-to-Point Communication on Heterogeneous Architectures},
    pdfauthor={Lockhart, Bienz, Gropp, Olson},     %
    colorlinks=true,       %
    linkcolor=green!50!black, %
    citecolor=blue,      %
    urlcolor=blue        %
}
\usepackage[capitalise,noabbrev]{cleveref}
\crefformat{equation}{(#2#1#3)}
\numberwithin{equation}{section}
\numberwithin{figure}{section}

\usepackage{ifthen}
\usepackage{tikz}
\usetikzlibrary{arrows}
\usetikzlibrary{calc,decorations.pathreplacing,shapes.misc}
\usetikzlibrary{patterns}
\usetikzlibrary{calc,positioning}
\usepackage{import}

\usepackage{siunitx} %

\def\BibTeX{{\rm B\kern-.05em{\sc i\kern-.025em b}\kern-.08em
    T\kern-.1667em\lower.7ex\hbox{E}\kern-.125emX}}

\newcommand{\Rn}{\mathbb{R}^n}

\newcommand{\Rmn}{\mathbb{R}^{m\times\ n}}

\newcommand{\ppn}{\texttt{ppn}}

\makeatletter
\def\ps@pprintTitle{%
  \let\@oddhead\@empty
  \let\@evenhead\@empty
  \def\@oddfoot{\reset@font\hfil\thepage\hfil}
  \let\@evenfoot\@oddfoot
}
\makeatother

\begin{document}

\begin{frontmatter}
\title{Characterizing the Performance of Node-Aware Strategies for Irregular Point-to-Point Communication on Heterogeneous Architectures}

\author[uiuc]{Shelby Lockhart\corref{cor1}}
\ead{sll2@illinois.edu}
\cortext[cor1]{Corresponding author}

\affiliation[uiuc]{organization={University of Illinois at Urbana-Champaign},
                   addressline={Department of Computer Science},
                   city={Urbana},
                   postcode={61801},
                   state={IL},
                   country={USA}}

\author[unm]{Amanda Bienz}
\ead{bienz@unm.edu}

\affiliation[unm]{organization={University of New Mexico},
                  addressline={Department of Computer Science},
                  city={Albuquerque},
                  postcode={87131},
                  state={NM},
                  country={USA}}

\author[uiuc]{William D. Gropp}
\ead{wgropp@illinois.edu}

\author[uiuc]{Luke N. Olson}
\ead{lukeo@illinois.edu}

\begin{abstract}
    Supercomputer architectures are trending toward higher
    computational throughput due to the inclusion of heterogeneous compute
    nodes. These multi-GPU nodes increase
    on-node computational efficiency, while also increasing
    the amount of data to be communicated and the number of potential
    data flow paths.
    In this work, we characterize the performance of irregular point-to-point
    communication with MPI on heterogeneous compute environments through
    performance modeling, demonstrating the limitations of standard
    communication strategies for both device-aware and staging-through-host
    communication techniques.
    Presented models suggest staging communicated data through host processes then
    using node-aware communication strategies for high inter-node message counts.
    Notably, the models also predict that node-aware communication
    utilizing all available CPU cores to communicate inter-node data leads to the
    most performant strategy when communicating with a high number of nodes.
    Model validation is provided via a case study of irregular point-to-point
    communication patterns in distributed sparse matrix-vector products.
    Importantly, we include a discussion on the implications model predictions
    have on communication strategy design for emerging supercomputer architectures.
\end{abstract}


\end{frontmatter}

\section{Introduction}

Modern parallel supercomputers exhibit increasingly higher computational throughput due to the
inclusion of multiple GPUs per node~---~see~\cref{ss:modern_arch}.
These GPUs operate on much higher
data volumes concurrently than previous CPU-only clusters,
yet the issue of communication bottlenecks persists and is exacerbated
in a multi-node--multi-GPU setting.  %
While the high computational intensity of modern supercomputers
is driving a new era of applications, the volume of data communicated between compute units has
also increased, creating new hurdles for data movement performance.

In this paper, we focus on irregular point-to-point communication,
which generates a performance bottleneck in parallel solvers and graph
algorithms due to the prevalence of sparse matrix operations and unstructured mesh
computations~\cite{mohiyuddin2009minimizing,smith2018improving}.
We aim to characterize the performance of various
irregular point-to-point communication strategies using MPI within
heterogeneous compute environments via performance modeling, which
suggests the extension of node-aware communication strategies for inter-CPU communication
(discussed in~\cref{ss:na}) onto heterogeneous architectures.

Node-aware communication schemes take advantage of the fact
that the performance of communication depends on
the relative location of communicating processes.
These schemes reduce communication times by exchanging more costly
data flow paths for cheaper ones when possible~\cite{Bienz_napspmv}.
While there are many potential paths for
data movement on heterogeneous architectures, we consider the communication
paths available via the MPI API and only consider device specific optimizations,
such as utilizing CUDA Multi-Process Service (MPS) to allow multiple MPI ranks to
copy data from a single GPU, for the purpose of comparison.

In~\cref{sec:modeling_params}, we present modeling parameters for all
potential data flow paths between CPUs and GPUs, which are then used within
performance models to predict the cost of various node-aware communication
schemes when implemented on heterogeneous architectures in~\cref{sec:na-hetero}.
Models are first validated via comparison against the performance of communication within
a sparse matrix-vector product, then
further modeling results are shown suggesting that for large message counts,
optimal performance is achieved when GPU data is staged through a host
process and split across multiple processes before communicating
through the network.

Furthermore,~\cref{sec:spmv} provides a study of the techniques modeled
in~\cref{sec:na-hetero} when applied to the irregular point-to-point communication
patterns in distributed sparse matrix-vector multiplication (SpMV) on heterogeneous
architectures, further validating model predictions.
Finally,~\cref{sec:conclusion} provides a discussion on the implications model predictions
and benchmark results have on the future of communication strategy design for emerging supercomputer architectures,
alongside a summary of the presented results.

The following provides a summary of paper contributions:
\begin{enumerate}
    \item performance models for node-aware
          communication on heterogeneous architectures~---~\cref{sec:modeling_params,sec:na-hetero};
    \item performance predictions for common irregular point-to-point communication scenarios using
          the developed models~---~\cref{ss:model_plots};
    \item benchmarks of irregular point-to-point communication patterns found within distributed sparse
          matrix-vector multiplication~---~\cref{sec:spmv}; and
    \item remarks on future communication design
          for emerging supercomputer architectures~---~\cref{sec:conclusion}.
\end{enumerate}

\section{Background}
\subsection{Modern Architectures}\label{ss:modern_arch}

Many current large-scale supercomputers consist of heterogeneous nodes
containing multiple GPUs connected to a single CPU per socket
with two sockets per node.
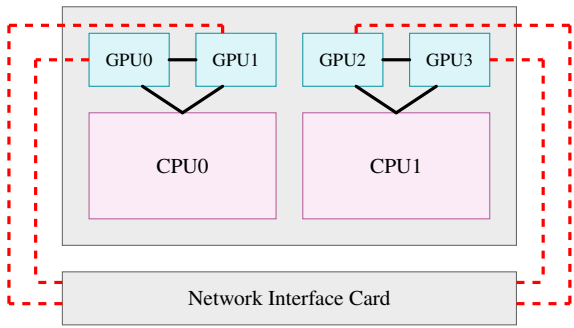
\begin{figure}[!ht]
    \small %
    \centering
    \definecolor{tab-blue}{HTML}{1f77b4}
\definecolor{tab-orange}{HTML}{ff7f0e}
\definecolor{tab-green}{HTML}{2ca02c}
\definecolor{tab-red}{HTML}{d62728}
\definecolor{tab-purple}{HTML}{9467bd}
\definecolor{tab-brown}{HTML}{8c564b}
\definecolor{tab-pink}{HTML}{e377c2}
\definecolor{tab-gray}{HTML}{7f7f7f}
\definecolor{tab-olive}{HTML}{bcbd22}
\definecolor{tab-cyan}{HTML}{17becf}

\newcommand{\heteronode}
{
  \begin{scope}[x=40pt,y=40pt]

  \draw[draw=tab-gray!80!black, fill=tab-gray!15] (0, 0) rectangle + (4.25, 0.5);
  \coordinate (nic) at (2.125, 0.25);
  \node[anchor=center] at (nic) {\footnotesize Network Interface Card};

  \draw[draw=tab-gray!80!black, fill=tab-gray!15] (0, 0.75) rectangle + (4.25, 2.25);

  \draw[draw=tab-pink!80!black, fill=tab-pink!15] (0.25, 1.0) rectangle  + (1.75, 1.0);
  \draw[draw=tab-pink!80!black, fill=tab-pink!15] (2.25, 1.0) rectangle  + (1.75, 1.0);
  \coordinate (cpu0) at (1.125, 1.5);
  \coordinate (cpu1) at (3.125, 1.5);
  \node[anchor=center] at (cpu0) {\footnotesize CPU0};
  \node[anchor=center] at (cpu1) {\footnotesize CPU1};

  \draw[draw=tab-cyan!80!black, fill=tab-cyan!15] (0.25, 2.25) rectangle  + (0.75, 0.5);
  \draw[draw=tab-cyan!80!black, fill=tab-cyan!15] (1.25, 2.25) rectangle  + (0.75, 0.5);
  \draw[draw=tab-cyan!80!black, fill=tab-cyan!15] (2.25, 2.25) rectangle  + (0.75, 0.5);
  \draw[draw=tab-cyan!80!black, fill=tab-cyan!15] (3.25, 2.25) rectangle  + (0.75, 0.5);
  \coordinate (gpu0) at (0.625, 2.5);
  \coordinate (gpu1) at (1.625, 2.5);
  \coordinate (gpu2) at (2.625, 2.5);
  \coordinate (gpu3) at (3.625, 2.5);
  \node[anchor=center] at (gpu0) {\scriptsize GPU0};
  \node[anchor=center] at (gpu1) {\scriptsize GPU1};
  \node[anchor=center] at (gpu2) {\scriptsize GPU2};
  \node[anchor=center] at (gpu3) {\scriptsize GPU3};

  \end{scope}
}

\begin{tikzpicture}[x=40pt,y=40pt]
  \begin{scope}[shift={(0pt, 0pt)}]
  \heteronode
  \tikzstyle{myarrow}=[>=latex,shorten >=5pt,shorten <=5pt,->,draw,thick,color=black]

  \draw[black, very thick, line cap=round] (1.0, 2.5) -- (1.25, 2.5);
  \draw[black, very thick, line cap=round] (3.0, 2.5) -- (3.25, 2.5);

  \draw[black, very thick, line cap=round] (0.75, 2.25) -- (1.125, 2.0);
  \draw[black, very thick, line cap=round] (1.5, 2.25) -- (1.125, 2.0);
  \draw[black, very thick, line cap=round] (2.75, 2.25) -- (3.125, 2.0);
  \draw[black, very thick, line cap=round] (3.5, 2.25) -- (3.125, 2.0);

  \draw[red, very thick, dashed] (0.25, 2.5) -- (-0.25, 2.5);
  \draw[red, very thick, dashed] (0, 0.4) -- (-0.25, 0.4);
  \draw[red, very thick, dashed] (-0.25, 2.5) -- (-0.25, 0.4);

  \draw[red, very thick, dashed] (1.5, 2.75) -- (1.5, 2.825);
  \draw[red, very thick, dashed] (0, 0.2) -- (-0.5, 0.2);
  \draw[red, very thick, dashed] (-0.5, 2.825) -- (-0.5, 0.2);
  \draw[red, very thick, dashed] (-0.5, 2.825) -- (1.5, 2.825);

  \draw[red, very thick, dashed] (2.75, 2.75) -- (2.75, 2.825);
  \draw[red, very thick, dashed] (4.25, 0.2) -- (4.75, 0.2);
  \draw[red, very thick, dashed] (4.75, 2.825) -- (4.75, 0.2);
  \draw[red, very thick, dashed] (4.75, 2.825) -- (2.75, 2.825);

  \draw[red, very thick, dashed] (4.0, 2.5) -- (4.5, 2.5);
  \draw[red, very thick, dashed] (4.25, 0.4) -- (4.5, 0.4);
  \draw[red, very thick, dashed] (4.5, 2.5) -- (4.5, 0.4);
  \end{scope}
\end{tikzpicture}

    \caption{Lassen compute node.}\label{fig:lassen_node}
\end{figure}
In the case of the Lassen supercomputer, each socket consists
of a single IBM Power9 CPU connected to two NVIDIA V100 GPUs~\cite{lassen}
(see~\cref{fig:lassen_node}),
while the Summit supercomputer has a single IBM Power9 CPU
connected to three NVIDIA V100 GPUs~\cite{summit}.
For both machines, each CPU has 20 available cores, CPUs and GPUs are connected
via NVLink, and nodes are connected via Mellanox EDR 100G
InfiniBand in a non-blocking fat tree topology.
Upcoming Department of Energy exascale machines, Frontier~\cite{frontier} and El Capitan\footnote{\url{https://www.llnl.gov/news/el-capitan-testbed-systems-rank-among-top-200-worlds-most-powerful-computers}}, will have nodes with a similar structure to those found in Lassen and Summit.
However, these compute nodes will consist of a single socket housing
an AMD EPYC CPU connected to four AMD Instinct 250X GPUs via AMD
Infinity Fabric with a Slingshot network. Additionally, the National Center for Supercomputing Applications (NCSA) will
boast more expansive compute nodes consisting of four to eight AMD A100 GPUs connected to a dual AMD 64-core 2.55 GHz Milan processor
per compute node in their upcoming system Delta\footnote{\url{https://www.ncsa.illinois.edu/research/project-highlights/delta}}.

Both current and future supercomputers boast heterogeneous architectures
with multiple paths for data movement between two GPUs.
Two connected GPUs either exchange data directly or
stage through the host CPU by first copying data to CPU memory,
then transferring data from the local CPU to the host CPU of the receiving GPU,
and finally copying received data to the destination GPU\@.
The process of staging data through the host CPU can be used for any
set of communicating GPUs independent of their relative locations.
However, device-aware data movement paradigms,
such as CUDA-aware MPI using GPUDirect~\cite{gpudirect} on Lassen,
remove the necessity of copying data to the host CPU and allow data
to be pulled directly from device memory, even in the case of inter-node
data transfers.
The addition of device-aware technologies increases the number of potential
data movement paths necessitating the use of robust performance modeling to
determine communication bottlenecks, as well as, design optimal
communication strategies.

\subsection{Modeling Data Movement}

Throughout this paper, we rely on the \textit{max-rate}
model as the basis for communication modeling~\cite{MaxRate}.
The max-rate model is an improvement to the standard postal model
of communication, accounting for injection limits into the network.
The traditional postal model estimates the cost of communicating a message
between two symmetric multiprocessing (SMP) nodes as
\begin{equation}\label{eq:postal}
    T = \alpha + \beta \cdot s
\end{equation}
where $\alpha$ is the latency, $\beta$ is the per-byte transfer cost, and
$s$ is the number of bytes being communicated.
The max-rate model adds parameters for injection-bandwidth limits and the
number of actively communicating processes, resulting in the following
time estimation,
\begin{equation}\label{eq:max_rate}
    T = \alpha \cdot m +
    \max\left(
              \frac{\ppn \cdot s}{R_N}, \frac{s}{R_b}
        \right)
\end{equation}
where $\alpha$ is again the latency, $m$ is the maximum number of messages sent by a
single process on a given node,
$s$ is the maximum number of bytes sent by a single process on a given SMP,
\ppn\ is the number of processes per node,
$R_N$ is the rate at which a network interface card (NIC) can inject data
into the network, and $R_b$ is the rate at which a process can transport data.
When $\ppn \cdot R_b < R_N$, this model reduces to the
postal model.

For inter-CPU communication, additional improvements are available to the
max-rate model within the context of irregular point-to-point communication.
Additional hardware and software overhead penalties are represented in the LogP model\cite{logP},
which is extended to include long message costs in the LogGP model\cite{logGP}.
The importance of modeling
queue search times and network contention on accurately predicting performance,
for example, is shown in~\cite{BienzEuroMPI}.
The additional parameters needed to account for contention and queue times motivates design decisions within the context
of node-aware communication techniques, highlighting trade-offs between
communication schemes.
Node-aware communication strategies are discussed in more
detail in~\cref{ss:na}.

The max-rate model also applies to inter-GPU communication~\cite{Bienz_modeling_hetero}.
Here, the noted difficulty in
reaching injection bandwidth limits with inter-GPU communication is due to the
low number of communicating GPUs per node. Additionally, for large message
counts, performance benefits are observed~\cite{Bienz_modeling_hetero}
when staging communication between GPUs through host CPUs.
In~\cref{sec:na-hetero}, the models for inter-GPU irregular
point-to-point communication are presented.

\subsection{Node-Aware Communication}\label{ss:na}

Node-aware communication techniques for irregular point-to-point
communication have been designed within
the context of sparse matrix-vector multiplication (SpMV) and
sparse matrix-matrix multiplication (SpMM)~\cite{Bienz_napspmv}.
Due to their low computational requirements, sparse matrix
operations often incur a large communication overhead when
performed in a parallel distributed setting,
highlighting the limitations of standard communication practices.

\begin{figure}[!ht]
    \small %
    \definecolor{tab-blue}{HTML}{1f77b4}
\definecolor{tab-orange}{HTML}{ff7f0e}
\definecolor{tab-green}{HTML}{2ca02c}
\definecolor{tab-red}{HTML}{d62728}
\definecolor{tab-purple}{HTML}{9467bd}
\definecolor{tab-brown}{HTML}{8c564b}
\definecolor{tab-pink}{HTML}{e377c2}
\definecolor{tab-gray}{HTML}{7f7f7f}
\definecolor{tab-olive}{HTML}{bcbd22}
\definecolor{tab-cyan}{HTML}{17becf}

\newcommand{\anode}
{
  \begin{scope}[x=28pt,y=28pt]
  \draw[draw=tab-gray!80!black, fill=tab-gray!15] (0,     0) rectangle +(2, 2);
  \draw[draw=tab-orange!80!black,fill=tab-orange!45]  (0.5, 1.5) circle (0.35 and .45);
  \draw[draw=tab-green!80!black, fill=tab-green!45]   (1.5, 1.5) circle (0.35 and .45);
  \draw[draw=tab-red!80!black,   fill=tab-red!45]     (0.5, 0.5) circle (0.35 and .45);
  \draw[draw=tab-blue!80!black,  fill=tab-blue!45]    (1.5, 0.5) circle (0.35 and .45);
  \coordinate (p0) at (0.5, 1.4);
  \coordinate (p1) at (1.5, 1.4);
  \coordinate (p2) at (0.5, 0.4);
  \coordinate (p3) at (1.5, 0.4);
  \node[anchor=center] at (p0) {\scriptsize P0};
  \node[anchor=center] at (p1) {\scriptsize P1};
  \node[anchor=center] at (p2) {\scriptsize P2};
  \node[anchor=center] at (p3) {\scriptsize P3};
  \node[anchor=south] at (1.0, 2.0) {Node 0};

  \begin{scope}[shift={(60pt,0pt)}]
  \draw[dashed,draw=tab-gray!80!black, fill=tab-gray!15] (0,     0) rectangle +(2, 2);
  \draw[dashed,draw=tab-orange!80!black,fill=tab-orange!45]  (0.5, 1.5) circle (0.35 and .45);
  \draw[dashed,draw=tab-green!80!black, fill=tab-green!45]   (1.5, 1.5) circle (0.35 and .45);
  \draw[dashed,draw=tab-red!80!black,   fill=tab-red!45]     (0.5, 0.5) circle (0.35 and .45);
  \draw[dashed,draw=tab-blue!80!black,  fill=tab-blue!45]    (1.5, 0.5) circle (0.35 and .45);
  \coordinate (p4) at (0.5, 1.4);
  \coordinate (p5) at (1.5, 1.4);
  \coordinate (p6) at (0.5, 0.4);
  \coordinate (p7) at (1.5, 0.4);
  \node[anchor=center] at (p4) {\scriptsize P4};
  \node[anchor=center] at (p5) {\scriptsize P5};
  \node[anchor=center] at (p6) {\scriptsize P6};
  \node[anchor=center] at (p7) {\scriptsize P7};
  \node[anchor=south] at (1.0, 2.0) {Node 1};
  \end{scope}
  \end{scope}
}

\begin{tikzpicture}[x=80pt,y=80pt]
  \begin{scope}[shift={(3pt, 15pt)}]
  \anode
  \tikzstyle{myarrow}=[>=latex,shorten >=5pt,shorten <=5pt,->,draw,thin,color=black]

  \path[myarrow] (p0) to (p6);
  \path[myarrow] (p1) to (p6);
  \path[myarrow] ([yshift=-10pt]p2.south) to (p6);
  \path[myarrow] (p3) to (p6);
  \path[myarrow] (p4) to (p6);
  \path[myarrow] (p5) to (p6);
  \path[myarrow] (p7) to (p6);

  \node at (60pt, -12.0pt) {Sending multiple messages};
  \node at (60pt, -22.0pt) {to Node 1};

  \end{scope}
  \begin{scope}[shift={(130pt, 15pt)}]
  \anode
  \tikzstyle{myarrow}=[>=latex,shorten >=5pt,shorten <=0pt,->,draw,thin,color=black]

  \foreach \j in {-4,0,4}{
    \draw[color=black!70, fill=tab-gray] ([xshift=\j pt,yshift=5pt]p1) rectangle +(0.04, 0.04);
  }
  \coordinate (myp) at ([xshift=9pt,yshift=6pt] p1);
  \path[myarrow] (myp) to[bend left] (p4);
  \path[myarrow] (myp) to[bend left] (p5);
  \path[myarrow] (myp) to[looseness=.7, bend right=45] (p6);
  \path[myarrow] (myp) to[bend right=75,in=90,out=-45] (p7);
  \node at (60pt, -12.0pt) {Sending duplicate data};
  \node at (60pt, -22.0pt) {to Node 1};

  \end{scope}
\end{tikzpicture}

    \caption{Standard communication.
    On the left, Node 0
    injects multiple messages into the network, all to P6 on Node 1.
    On the right, P1 sends all highlighted data to multiple processes on Node 1, leading
    to redundant messages.}\label{fig:standard_comm}
\end{figure}
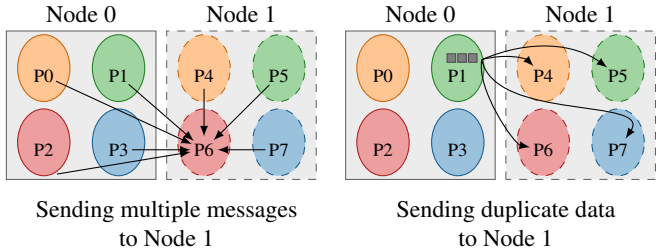

There are two redundancies that occur with standard communication, namely: a message redundancy and a data redundancy,
illustrated in~\cref{fig:standard_comm}.
First, each node injects many messages into the network;
for example, some nodes send multiple messages to a
single process on the destination node creating message redundancy.
Second, processes send their local data to
every destination process, independent of whether they had sent
the same local data to another process on the same node; hence a
redundancy in data being sent through the network.
The majority of node-aware communication work has been done
within the context of CPU to CPU communication with a subset of
this work later replicated for GPU to GPU communication.
There are three types of node-aware communication for CPU to CPU
communication, each eliminating all or some of the redundancies
introduced by standard communication.

\subsubsection{3-Step}

3-Step node-aware communication, first introduced
in~\cite{Bienz_napspmv}, eliminates both redundancies in
standard communication by gathering all necessary data to be sent off-node in
a single buffer. Pairing all processes with a receiving process on distinct
nodes ensures efficiency of the method by making sure every process remains active
throughout the communication scheme.
\begin{figure}[!ht]
    \centering
    \small %
    \definecolor{tab-blue}{HTML}{1f77b4}
\definecolor{tab-orange}{HTML}{ff7f0e}
\definecolor{tab-green}{HTML}{2ca02c}
\definecolor{tab-red}{HTML}{d62728}
\definecolor{tab-purple}{HTML}{9467bd}
\definecolor{tab-brown}{HTML}{8c564b}
\definecolor{tab-pink}{HTML}{e377c2}
\definecolor{tab-gray}{HTML}{7f7f7f}
\definecolor{tab-olive}{HTML}{bcbd22}
\definecolor{tab-cyan}{HTML}{17becf}

\newcommand{\anode}
{
  \begin{scope}[x=28pt,y=28pt]
  \draw[draw=tab-gray!80!black, fill=tab-gray!15] (0,     0) rectangle +(2, 2);
  \draw[draw=tab-orange!80!black,fill=tab-orange!45]  (0.5, 1.5) circle (0.35 and .45);
  \draw[draw=tab-green!80!black, fill=tab-green!45]   (1.5, 1.5) circle (0.35 and .45);
  \draw[draw=tab-red!80!black,   fill=tab-red!45]     (0.5, 0.5) circle (0.35 and .45);
  \draw[draw=tab-blue!80!black,  fill=tab-blue!45]    (1.5, 0.5) circle (0.35 and .45);
  \coordinate (p0) at (0.5, 1.4);
  \coordinate (p1) at (1.5, 1.4);
  \coordinate (p2) at (0.5, 0.4);
  \coordinate (p3) at (1.5, 0.4);
  \node[anchor=center] at (p0) {\scriptsize P0};
  \node[anchor=center] at (p1) {\scriptsize P1};
  \node[anchor=center] at (p2) {\scriptsize P2};
  \node[anchor=center] at (p3) {\scriptsize P3};
  \node[anchor=south] at (1.0, 2.0) {Node 0};

  \begin{scope}[shift={(60pt,0pt)}]
  \draw[dashed,draw=tab-gray!80!black, fill=tab-gray!15] (0,     0) rectangle +(2, 2);
  \draw[dashed,draw=tab-orange!80!black,fill=tab-orange!45]  (0.5, 1.5) circle (0.35 and .45);
  \draw[dashed,draw=tab-green!80!black, fill=tab-green!45]   (1.5, 1.5) circle (0.35 and .45);
  \draw[dashed,draw=tab-red!80!black,   fill=tab-red!45]     (0.5, 0.5) circle (0.35 and .45);
  \draw[dashed,draw=tab-blue!80!black,  fill=tab-blue!45]    (1.5, 0.5) circle (0.35 and .45);
  \coordinate (p4) at (0.5, 1.4);
  \coordinate (p5) at (1.5, 1.4);
  \coordinate (p6) at (0.5, 0.4);
  \coordinate (p7) at (1.5, 0.4);
  \node[anchor=center] at (p4) {\scriptsize P4};
  \node[anchor=center] at (p5) {\scriptsize P5};
  \node[anchor=center] at (p6) {\scriptsize P6};
  \node[anchor=center] at (p7) {\scriptsize P7};
  \node[anchor=south] at (1.0, 2.0) {Node 1};
  \end{scope}
  \end{scope}
}

\tikzset{
  doublearrowp/.style args={#1 and #2}{
    round cap-latex,line width=1pt,#1,shorten <=5pt,
    postaction={draw,round cap-latex,#2,line width=1pt/3,shorten <=5.5pt,shorten >=1.5pt},
  },
  doublearrowq/.style args={#1 and #2}{
    round cap-latex,line width=1pt,#1, shorten <=2pt, shorten >=5pt,
    postaction={draw,round cap-latex,#2,line width=1pt/3,shorten <=2.5pt,shorten >=6.5pt},
  },
  doublearrowz/.style args={#1 and #2}{
    round cap-latex,line width=1pt,#1,shorten <=2pt, shorten >=2pt,
    postaction={draw,round cap-latex,#2,line width=1pt/3,shorten <=2.5pt,shorten >=3.5pt},
  },
  doublearrowp2/.style args={#1}{
    round cap-latex,line width=1pt,#1,shorten <=5pt,
  },
  doublearrowq2/.style args={#1}{
    round cap-latex,line width=1pt,#1, shorten <=2pt, shorten >=5pt,
  },
  doublearrowz2/.style args={#1}{
    round cap-latex,line width=1pt,#1,shorten <=2pt, shorten >=5pt,
  },
}
\begin{tikzpicture}[x=80pt,y=80pt]
  \begin{scope}[shift={(3pt, 5pt)}]
  \newcommand{\somenode}[3]{
  \begin{scope}
  \anode

  \foreach \j in {-4,0,4}{
    \draw[color=black!70, fill=tab-gray] ([xshift=\j pt,yshift=5pt]p0) rectangle +(0.04, 0.04);
  }
  \coordinate (myp) at ([xshift=9pt,yshift=6pt] p0);
  \draw[doublearrowp2=#1] (p1) to (myp);
  \draw[doublearrowp2=#1] (p2) to (myp);
  \draw[doublearrowp2=#1] (p3) to (myp);

  \foreach \j in {-4,0,4}{
    \draw[color=black!70, fill=tab-gray] ([xshift=\j pt,yshift=5pt]p7) rectangle +(0.04, 0.04);
  }
  \coordinate (myq) at ([xshift=-4pt,yshift=6pt] p7);
  \draw[doublearrowq2=#2] (myq) to (p4);
  \draw[doublearrowq2=#2] (myq) to (p5);
  \draw[doublearrowq2=#2] (myq) to (p6);

  \foreach \j in {-4,0,4}{
    \draw[color=black!70, fill=tab-gray] ([xshift=\j pt,yshift=5pt]p7) rectangle +(0.04, 0.04);
  }

  \draw[doublearrowz2=#3] (myp) to[out=80, in=160] (myq);
    \end{scope}
  }

  \begin{scope}[xshift=20pt,yshift=150pt]
    \node[rotate=90] at (-10pt,25pt) {Step 1};
    \somenode{dotted}{solid}{solid}
  \end{scope}
  \begin{scope}[xshift=65pt,yshift=75pt]
    \node[rotate=90] at (-10pt,25pt) {Step 2};
    \somenode{solid}{solid}{dotted}
  \end{scope}
  \begin{scope}[xshift=125pt,yshift=0pt]
    \node[rotate=90] at (-10pt,25pt) {Step 3};
    \somenode{solid}{dotted}{solid}
  \end{scope}

  \end{scope}
\end{tikzpicture}

    \caption{3-Step node-aware.
    In Step 1, all data on
    Node 0 that needs to be sent to Node 1 is collected in a buffer
    on P0, the process paired to send and receive from Node 1.
    In Step 2, P0 sends this buffer from Node 0 to P7,
    the receiving process on Node 1. In Step 3, P7
    redistributes the data to the correct receiving processes on Node 1.
    Dotted lines,~\protect\tikz{\protect\draw[dotted,thick] (0,0)--(0.5,0);}, %
    depict the action performed in each step.}\label{fig:3step}
\end{figure}
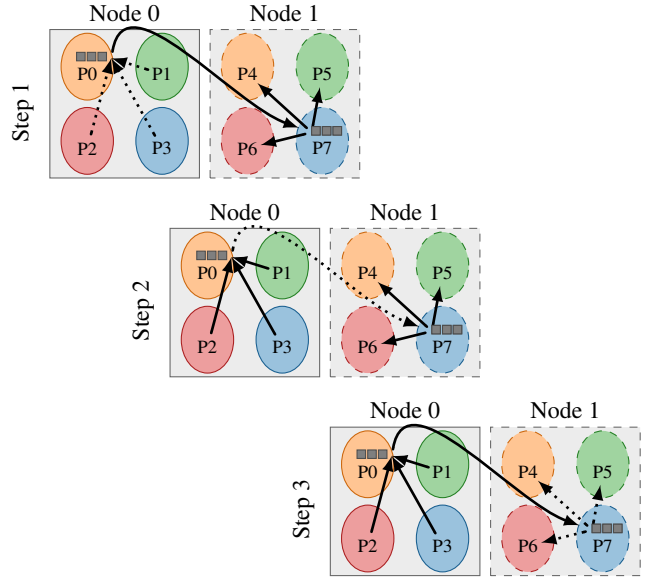
First, all messages sent to a separate node are
gathered in a buffer by the single process associated with the node.
Secondly, this process sends the data buffer to the paired process on the
receiving node. Thirdly, the paired process on the receiving node redistributes
the data to the correct destination processes on-node. An example of these steps
is outlined in~\cref{fig:3step}.

As noted in~\cite{Bienz_napspmv}, the method can
be extended to include further breakdown of data exchanges to include
intra-socket data communication before the intra-node communication phase.
However, we expect minimal performance benefits in extending the communication
strategy throughout the entire node hierarchy for CPU to CPU communication.
Instead, this strategy is adopted for GPU to GPU communication in~\cite{petascale_xct},
where the full hierarchy of the node is utilized to achieve optimal performance
due to the fast data transfer rates of socket-level GPU interconnects on Summit~\cite{summit}.
In addition, recent work on utilizing neighborhood
collectives in conjunction with the 3-Step node-aware communication strategy further
reduces communication overhead in sparse solvers~\cite{Bienz_siampp22}.

\subsubsection{2-Step}

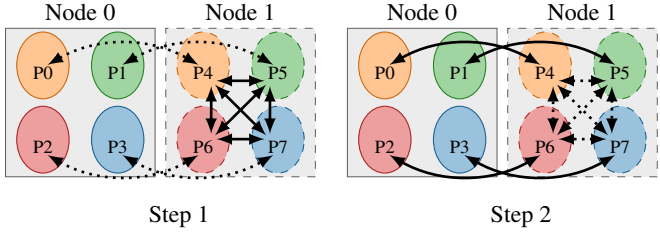
\begin{figure}[!htb]
    \centering
    \small %
    \definecolor{tab-blue}{HTML}{1f77b4}
\definecolor{tab-orange}{HTML}{ff7f0e}
\definecolor{tab-green}{HTML}{2ca02c}
\definecolor{tab-red}{HTML}{d62728}
\definecolor{tab-purple}{HTML}{9467bd}
\definecolor{tab-brown}{HTML}{8c564b}
\definecolor{tab-pink}{HTML}{e377c2}
\definecolor{tab-gray}{HTML}{7f7f7f}
\definecolor{tab-olive}{HTML}{bcbd22}
\definecolor{tab-cyan}{HTML}{17becf}

\newcommand{\anode}
{
  \begin{scope}[x=28pt,y=28pt]
  \draw[draw=tab-gray!80!black, fill=tab-gray!15] (0,     0) rectangle +(2, 2);
  \draw[draw=tab-orange!80!black,fill=tab-orange!45]  (0.5, 1.5) circle (0.35 and .45);
  \draw[draw=tab-green!80!black, fill=tab-green!45]   (1.5, 1.5) circle (0.35 and .45);
  \draw[draw=tab-red!80!black,   fill=tab-red!45]     (0.5, 0.5) circle (0.35 and .45);
  \draw[draw=tab-blue!80!black,  fill=tab-blue!45]    (1.5, 0.5) circle (0.35 and .45);
  \coordinate (p0) at (0.5, 1.4);
  \coordinate (p1) at (1.5, 1.4);
  \coordinate (p2) at (0.5, 0.4);
  \coordinate (p3) at (1.5, 0.4);
  \node[anchor=center] at (p0) {\scriptsize P0};
  \node[anchor=center] at (p1) {\scriptsize P1};
  \node[anchor=center] at (p2) {\scriptsize P2};
  \node[anchor=center] at (p3) {\scriptsize P3};
  \node[anchor=south] at (1.0, 2.0) {Node 0};

  \begin{scope}[shift={(60pt,0pt)}]
  \draw[dashed,draw=tab-gray!80!black, fill=tab-gray!15] (0,     0) rectangle +(2, 2);
  \draw[dashed,draw=tab-orange!80!black,fill=tab-orange!45]  (0.5, 1.5) circle (0.35 and .45);
  \draw[dashed,draw=tab-green!80!black, fill=tab-green!45]   (1.5, 1.5) circle (0.35 and .45);
  \draw[dashed,draw=tab-red!80!black,   fill=tab-red!45]     (0.5, 0.5) circle (0.35 and .45);
  \draw[dashed,draw=tab-blue!80!black,  fill=tab-blue!45]    (1.5, 0.5) circle (0.35 and .45);
  \coordinate (p4) at (0.5, 1.4);
  \coordinate (p5) at (1.5, 1.4);
  \coordinate (p6) at (0.5, 0.4);
  \coordinate (p7) at (1.5, 0.4);
  \node[anchor=center] at (p4) {\scriptsize P4};
  \node[anchor=center] at (p5) {\scriptsize P5};
  \node[anchor=center] at (p6) {\scriptsize P6};
  \node[anchor=center] at (p7) {\scriptsize P7};
  \node[anchor=south] at (1.0, 2.0) {Node 1};
  \end{scope}
  \end{scope}
}

\tikzset{
  doublearrowp/.style args={#1 and #2}{
    latex-latex,line width=1pt,#1,                   shorten <=2pt,  shorten >=2pt,
    postaction={draw,latex-latex,#2,line width=1pt/3,shorten <=3.5pt,shorten >=3.5pt},
  },
  doublearrowq/.style args={#1 and #2}{
    latex-latex,line width=1pt,#1,                   shorten <=2pt,  shorten >=2pt,
    postaction={draw,latex-latex,#2,line width=1pt/3,shorten <=3.5pt,shorten >=3.5pt},
  },
  doublearrowp2/.style args={#1}{
    latex-latex,line width=1pt,#1,                   shorten <=2pt,  shorten >=2pt,
  },
  doublearrowq2/.style args={#1}{
    latex-latex,line width=1pt,#1,                   shorten <=2pt,  shorten >=2pt,
  },
}
\begin{tikzpicture}[x=80pt,y=80pt]
  \begin{scope}[shift={(3pt, 25pt)}]
  \newcommand{\somenode}[2]{
  \begin{scope}
  \anode

  \draw[doublearrowp2=#1] ([xshift= 3pt, yshift=-3pt]p4) to ([xshift=-3pt, yshift=-3pt]p5);
  \draw[doublearrowp2=#1] ([xshift=-3pt, yshift=-3pt]p5) to ([xshift=-3pt, yshift= 3pt]p7);
  \draw[doublearrowp2=#1] ([xshift=-3pt, yshift= 3pt]p7) to ([xshift= 3pt, yshift= 3pt]p6);
  \draw[doublearrowp2=#1] ([xshift= 3pt, yshift= 3pt]p6) to ([xshift= 3pt, yshift=-3pt]p4);
  \draw[doublearrowp2=#1] ([xshift= 3pt, yshift=-3pt]p4) to ([xshift=-3pt, yshift= 3pt]p7);
  \draw[doublearrowp2=#1] ([xshift=-3pt, yshift=-3pt]p5) to ([xshift= 3pt, yshift= 3pt]p6);

  \draw[doublearrowq2=#2, bend left]  ([yshift=2pt]p0) to ([yshift=2pt]p4);
  \draw[doublearrowq2=#2, bend left]  ([yshift=2pt]p1) to ([yshift=2pt]p5);
  \draw[doublearrowq2=#2, bend right] ([yshift=-2pt]p2) to ([yshift=-2pt]p6);
  \draw[doublearrowq2=#2, bend right] ([yshift=-2pt]p3) to ([yshift=-2pt]p7);

  \end{scope}
  }

  \begin{scope}[xshift=0pt,yshift=0pt]
    \node at (65pt,-15pt) {Step 1};
    \somenode{solid}{dotted}
  \end{scope}
  \begin{scope}[xshift=128pt,yshift=0pt]
    \node at (65pt,-15pt) {Step 2};
    \somenode{dotted}{solid}
  \end{scope}

  \end{scope}
\end{tikzpicture}

    \caption{2-Step node-aware.
    Each process on Node 0
    is paired with a receiving process on Node 1. In Step 1, each
    process on Node 0 sends the data needed by any process on Node 1
    to its paired process on Node 1. Here, P0 is sending to P4, P1
    to P5, P2 to P6, and P3 to P7. In Step 2, each process on Node 1
    redistributes the data received from Node 0 to the destination
    on Node 1.
    Dotted lines,~\protect\tikz{\protect\draw[dotted,thick] (0,0)--(0.5,0);}, %
    depict the action performed in each step.}\label{fig:2step}
\end{figure}

When communicating high data volumes between nodes, 3-Step
communication can see limitations as the single buffer communicating
data grows extremely large, thus motivating a 2-Step node-aware technique as
in~\cite{Bienz_napamg}.
The 2-Step technique eliminates the redundancy of sending
duplicate data through the network, but does not reduce
the redundancy of multiple messages being sent between nodes.
In 2-Step, \textit{each}
process exchanges information
needed by the receiving node with their paired process directly,
followed by the receiving node redistributing the
messages on-node, as shown in~\cref{fig:2step}.
Overall, the total number of bytes communicated
with 3-Step and 2-Step communication techniques is the same,
but the number and size of inter-node messages differs.

\subsubsection{Split}
\begin{figure}[!ht]
    \centering
    \includegraphics[width=\columnwidth]{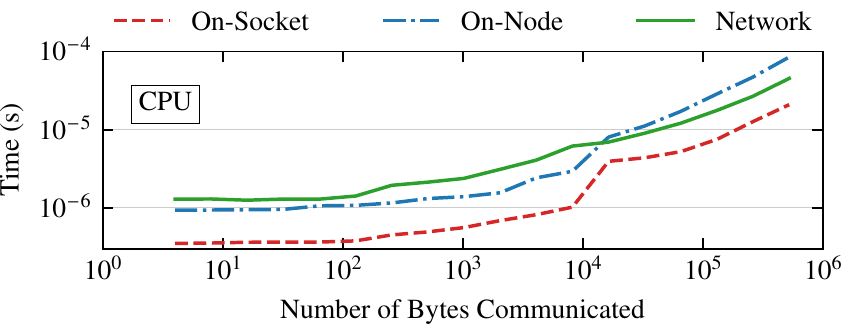}
    \caption{The amount of time required to send data between two processes
    distinguishing between where the two processes are physically located
    on the same socket, the same node and separate sockets, and separate nodes
    requiring network communication.}\label{fig:cpu_ping_pong}
\end{figure}

\begin{figure}[!ht]
    \centering
    \includegraphics[width=\columnwidth]{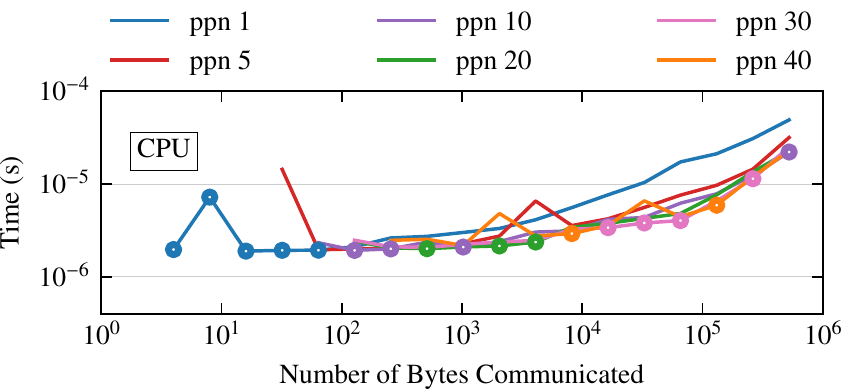}
    \caption{The amount of time required to send various amounts of data between two
    distinct nodes when splitting the data across ppn processes
    per node. Minimum times circled.}\label{fig:cpu_node_pong}
\end{figure}

3-Step and 2-Step communication
show a drastic difference in
performance in communicating on-node versus inter-node messages~\cite{Bienz_napspmv},
particularly on more traditional networks, e.g., the now retired BlueWaters system.
Yet this is not always the case for more recent interconnects, such as on
Lassen, which shows varying performance for inter-node versus intra-node
communication depending on the amount of data being communicated~---~see~\cref{fig:cpu_ping_pong}.
In addition to network communication being faster than on-node communication
for large message sizes, the CPUs used in current supercomputers have high
numbers of cores (for example, the IBM Power9 has 40 available cores on Lassen, and the Delta system has 64 available cores
on each AMD Milan processor),
making splitting large data volumes across all available cores more
performant than when the entire data volume is sent by a single process~---~see~\cref{fig:cpu_node_pong}.

Split communication, as introduced in~\cite{Lockhart_ecg_optna},
addresses the variable performance of 3-Step and 2-Step
node-aware communication on modern supercomputers. This communication
technique balances the performance of multi-step communication by splitting
the communicated inter-node data into messages of size
\texttt{message\_cap}, followed by a distribution across
some number of on-node processes before being injected into the
network.
Pseudo-code of the setup is provided in~\cref{alg:split_setup}
with communication parameters defined in~\cref{tab:split_parameters}.
Here, we detail the operations summarized in~\cref{alg:split_setup}.
\newcommand{\Split}{\textbf{Split}\xspace}
\newcommand{\Conglomerate}{\textbf{Conglomerate}\xspace}
\newcommand{\Create}{\textbf{Create}\xspace}
\newcommand{\Set}{\textbf{Set}\xspace}
\begin{vAlgorithm*}[!ht]{0.7\textwidth}{0.1in}
  \DontPrintSemicolon%
  \textbf{Input}:  \texttt{l\_recv}      \tcc*{list of messages to receive}
  \hspace{2.73em}  \texttt{comm}         \tcc*{world communicator}
  \hspace{2.73em}  \texttt{message\_cap} \tcc*{user-defined message cap size}
  \textbf{Output}: \texttt{local\_comm}  \tcc*{on-node subcommunicator}
  \hspace{3.45em}  \texttt{local\_Rcomm} \tcc*{redistribution subcommunicator}
  \hspace{3.45em}  \texttt{global\_comm} \tcc*{off-node subcommunicator}
  \hspace{3.45em}  \texttt{local\_Scomm} \tcc*{on-node sending subcommunicator} %
  \Split messages by origin, off-node and on-node\\
  \texttt{local\_comm} $\gets$ \textbf{Create} on-node communicator\\
  \Split off-node messages by node\\
  \Set parameters in~\cref{tab:split_parameters}\\
  \If{\emph{\texttt{max\_IN\_recv\_size}} $<$ \emph{\texttt{message\_cap}}}{
    \Conglomerate all inter-node receives by node
  }
  \Else{
    \If{$\frac{\text{\emph{\texttt{total\_IN\_recv\_vol}}}}{\text{\emph{\texttt{message\_cap}}}} > \text{\emph{\texttt{PPN}}}$ \emph{\&}
        \emph{\texttt{num\_IN\_nodes}} $<$ \emph{\texttt{PPN}}}{
      \Set $\texttt{message\_cap} = \lceil\frac{\texttt{total\_IN\_recv\_vol}}{\texttt{PPN}}\rceil$
    }
    \Split inter-node receives to max size \texttt{message\_cap}\\
  }
  \Set on-node receive order (descending by size) \\
  \texttt{local\_Rcomm} $\gets$ \Create redistribution communicator (receive)\\
  \texttt{global\_comm} $\gets$ \Create inter-node communicator\\
  \texttt{local\_Scomm} $\gets$ \Create redistribution communicator (send)\\
  \caption{Setup for split communication.}\label{alg:split_setup}
\end{vAlgorithm*}
\begin{table*}[t!]
    \renewcommand{\arraystretch}{1.1}
    \begin{center}
    \begin{tabular}{rl}
      \toprule
      Parameter & Description \\
      \midrule
      \texttt{message\_cap} & maximum message size when splitting communicated inter-nodal data volumes\\
      \texttt{total\_IN\_recv\_vol} & total amount of data being received by this node from any other
              node in Bytes\\
      \texttt{max\_IN\_recv\_size} & maximum amount of data being received from a single other node
              in Bytes\\
      \texttt{num\_IN\_nodes} & number of nodes from which this node is receiving any messages\\
      \texttt{PPN} & processes per node\\
      \bottomrule
    \end{tabular}
    \end{center}
    \caption{Split communication parameters.}\label{tab:split_parameters}
\end{table*}
\begin{vAlgorithm}[!ht]{\linewidth}{0.1in}
  \DontPrintSemicolon%
  Perform \texttt{local\_comm} communication. \\
  Perform \texttt{local\_Scomm} data redistribution.\label{alg:scomm} \\
  Perform \texttt{global\_comm} inter-node communication.\label{alg:gcomm} \\
  Perform \texttt{local\_Rcomm} data redistribution.\label{alg:rcomm} \\
  \caption{Split communication.}\label{alg:split_comm}
\end{vAlgorithm}

\begin{description}
  \item[Line 8] The algorithm begins by splitting inter-node messages by their origin node (on-node or off-node).
  \item[Line 9] A \emph{local} communicator is created for exchanging all messages with origin on-node.
  \item[Line 10] All messages with origin off-node are split into lists according to their origin node.
  \item[Lines 11] Parameters, such as the number of nodes from which this node receives, the \emph{maximum}
        amount of data being received from a single other node, and the \emph{total} amount of data being
        received from any node by this node, are determined.
  \item[Lines 12--17] In this block of the algorithm, the appropriate\\
        \texttt{message\_cap} is determined.
  \begin{description}
        \item[Lines 12--13] First, the maximum amount of data being received from any node is checked
            to determine if it is smaller than the user provided \texttt{message\_cap}.
            When this occurs, every node's data should be sent in a single message.
        \item[Lines 14--17] Otherwise, if the total inter-node data volume being communicated divided by
            the provided \texttt{ message\_cap} is greater than the active number of processes per node,
            then the \texttt{message\_cap} is increased to be the total inter-node data volume divided by
            the number of on-node processes.
  \end{description}
  \item[Line 18] On-node processes are assigned inter-node messages to receive in descending order of size,
        starting with local rank 0. Inter-node messages to be sent are assigned in the reverse order
        starting with local rank \texttt{PPN-1}. This in combination with the message splitting ensures that
        all processes are active during communication.
  \item[Line 19] A \emph{local} communicator is created for redistributing all received inter-node data to
        its final destination processes on-node.
  \item[Line 20] A \emph{global} communicator is created for exchanging inter-node messages based on send
        and receive message assignment in \textbf{Line 18}.
  \item[Line 21] A \emph{local} communicator is created for redistributing all inter-node data to be sent by
        this node to the local processes responsible for sending the inter-node messages.
\end{description}

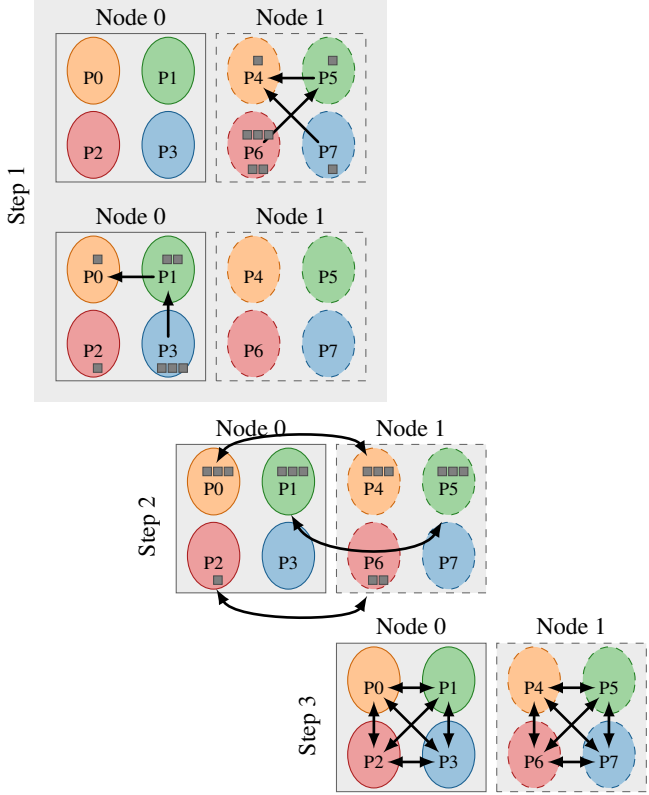
\begin{figure}[!ht]
  \centering
  \small %
  \definecolor{tab-blue}{HTML}{1f77b4}
\definecolor{tab-orange}{HTML}{ff7f0e}
\definecolor{tab-green}{HTML}{2ca02c}
\definecolor{tab-red}{HTML}{d62728}
\definecolor{tab-purple}{HTML}{9467bd}
\definecolor{tab-brown}{HTML}{8c564b}
\definecolor{tab-pink}{HTML}{e377c2}
\definecolor{tab-gray}{HTML}{7f7f7f}
\definecolor{tab-olive}{HTML}{bcbd22}
\definecolor{tab-cyan}{HTML}{17becf}

\newcommand{\anode}
{
  \begin{scope}[x=28pt,y=28pt]
  \draw[draw=tab-gray!80!black, fill=tab-gray!15] (0,     0) rectangle +(2, 2);
  \draw[draw=tab-orange!80!black,fill=tab-orange!45]  (0.5, 1.5) circle (0.35 and .45);
  \draw[draw=tab-green!80!black, fill=tab-green!45]   (1.5, 1.5) circle (0.35 and .45);
  \draw[draw=tab-red!80!black,   fill=tab-red!45]     (0.5, 0.5) circle (0.35 and .45);
  \draw[draw=tab-blue!80!black,  fill=tab-blue!45]    (1.5, 0.5) circle (0.35 and .45);
  \coordinate (p0) at (0.5, 1.4);
  \coordinate (p1) at (1.5, 1.4);
  \coordinate (p2) at (0.5, 0.4);
  \coordinate (p3) at (1.5, 0.4);
  \node[anchor=center] at (p0) {\scriptsize P0};
  \node[anchor=center] at (p1) {\scriptsize P1};
  \node[anchor=center] at (p2) {\scriptsize P2};
  \node[anchor=center] at (p3) {\scriptsize P3};
  \node[anchor=south] at (1.0, 2.0) {Node 0};

  \begin{scope}[shift={(60pt,0pt)}]
  \draw[dashed,draw=tab-gray!80!black, fill=tab-gray!15] (0,     0) rectangle +(2, 2);
  \draw[dashed,draw=tab-orange!80!black,fill=tab-orange!45]  (0.5, 1.5) circle (0.35 and .45);
  \draw[dashed,draw=tab-green!80!black, fill=tab-green!45]   (1.5, 1.5) circle (0.35 and .45);
  \draw[dashed,draw=tab-red!80!black,   fill=tab-red!45]     (0.5, 0.5) circle (0.35 and .45);
  \draw[dashed,draw=tab-blue!80!black,  fill=tab-blue!45]    (1.5, 0.5) circle (0.35 and .45);
  \coordinate (p4) at (0.5, 1.4);
  \coordinate (p5) at (1.5, 1.4);
  \coordinate (p6) at (0.5, 0.4);
  \coordinate (p7) at (1.5, 0.4);
  \node[anchor=center] at (p4) {\scriptsize P4};
  \node[anchor=center] at (p5) {\scriptsize P5};
  \node[anchor=center] at (p6) {\scriptsize P6};
  \node[anchor=center] at (p7) {\scriptsize P7};
  \node[anchor=south] at (1.0, 2.0) {Node 1};
  \end{scope}
  \end{scope}
}

\tikzset{
  doublearrowp/.style args={#1 and #2}{
    round cap-latex,line width=1pt,#1,shorten <=5pt,
    postaction={draw,round cap-latex,#2,line width=1pt/3,shorten <=5.5pt,shorten >=1.5pt},
  },
  doublearrowq/.style args={#1 and #2}{
    round cap-latex,line width=1pt,#1, shorten <=2pt, shorten >=5pt,
    postaction={draw,round cap-latex,#2,line width=1pt/3,shorten <=2.5pt,shorten >=6.5pt},
  },
  doublearrowz/.style args={#1 and #2}{
    round cap-latex,line width=1pt,#1,shorten <=2pt, shorten >=2pt,
    postaction={draw,round cap-latex,#2,line width=1pt/3,shorten <=2.5pt,shorten >=3.5pt},
  },
  doublearrowp2/.style args={#1}{
    round cap-latex,line width=1pt,#1,shorten <=5pt, shorten >=5pt,
  },
  doublearrowq2/.style args={#1}{
    round cap-latex,line width=1pt,#1, shorten <=2pt, shorten >=5pt,
  },
  doublearrowz2/.style args={#1}{
    round cap-latex,line width=1pt,#1,shorten <=5pt, shorten >=5pt,
  },
}
\begin{tikzpicture}[x=80pt,y=80pt]
  \begin{scope}[shift={(5pt, 5pt)}]
  \newcommand{\drawdata}[3]{
    \ifthenelse{\equal{#3}{above}}
    {
     \pgfmathsetmacro\zz{5pt}
    }
    {
     \pgfmathsetmacro\zz{-8pt}
    }

    \pgfmathsetmacro\m{#1-1}
    \foreach \j in {0,...,\m}{
      \draw[color=black!70, fill=tab-gray] ([xshift=-\m*2+4*\j pt,yshift=\zz]#2) rectangle +(0.04, 0.04);
    }
  }
  \begin{scope}[xshift=20pt,yshift=155pt]
    \fill[tab-gray!15] (-0.1,-0.1) rectangle +(1.65,1.9);
    \anode
    \draw[doublearrowp2=black] (p1) to (p0);
    \draw[doublearrowp2=black] (p3) to (p1);
    \drawdata{1}{p0}{above}
    \drawdata{2}{p1}{above}
    \drawdata{1}{p2}{below}
    \drawdata{3}{p3}{below}
    \begin{scope}[xshift=0pt,yshift=75pt]
    \anode
    \draw[doublearrowp2=black] (p5) to (p4);
    \draw[doublearrowp2=black] (p7) to (p4);
    \draw[doublearrowp2=black] (p6) to (p5);
    \drawdata{1}{p4}{above}
    \drawdata{1}{p5}{above}
    \drawdata{1}{p7}{below}
    \drawdata{3}{p6}{above}
    \drawdata{2}{p6}{below}
    \end{scope}
    \node[rotate=90] at (-14pt,70pt) {Step 1};
  \end{scope}
  \begin{scope}[xshift=65pt,yshift=75pt]
    \node[rotate=90] at (-10pt,25pt) {Step 2};
    \anode
    \draw[doublearrowq2=black, <->, >=latex] ([yshift=8pt]p0) to[looseness=0.7,out=60, in=120] ([yshift=8pt]p4);

    \draw[doublearrowq2=black, <->, >=latex]  ([yshift=-8pt]p2) to[looseness=0.7,out=-60, in=-120] ([yshift=-8pt]p6);
    \draw[doublearrowq2=black, <->, >=latex] ([yshift=-8pt]p1) to[looseness=0.9,out=-60, in=-120] ([yshift=-8pt]p5);

    \drawdata{3}{p0}{above}
    \drawdata{3}{p1}{above}
    \drawdata{1}{p2}{below}

    \drawdata{3}{p4}{above}
    \drawdata{3}{p5}{above}
    \drawdata{2}{p6}{below}
  \end{scope}
  \begin{scope}[xshift=125pt,yshift=0pt]
    \node[rotate=90] at (-10pt,25pt) {Step 3};
    \anode
    \draw[doublearrowz2=black, <->, >=latex] (p0) -- (p1);
    \draw[doublearrowz2=black, <->, >=latex] (p1) -- (p3);
    \draw[doublearrowz2=black, <->, >=latex] (p3) -- (p2);
    \draw[doublearrowz2=black, <->, >=latex] (p2) -- (p0);
    \draw[doublearrowz2=black, <->, >=latex] (p0) -- (p3);
    \draw[doublearrowz2=black, <->, >=latex] (p2) -- (p1);

    \draw[doublearrowz2=black, <->, >=latex] (p4) -- (p5);
    \draw[doublearrowz2=black, <->, >=latex] (p5) -- (p7);
    \draw[doublearrowz2=black, <->, >=latex] (p7) -- (p6);
    \draw[doublearrowz2=black, <->, >=latex] (p6) -- (p4);
    \draw[doublearrowz2=black, <->, >=latex] (p4) -- (p7);
    \draw[doublearrowz2=black, <->, >=latex] (p5) -- (p6);
  \end{scope}

  \end{scope}
\end{tikzpicture}

  \caption{Split node-aware. Here, data is communicated
  between two distinct nodes: Node 0 and Node 1, each
  with 4 local processes, denoted P\#.
  In Step 1, each node conglomerates small messages to be sent off-node,
  splits messages based on a message cap of 3, and retains messages approximately
  the size of the message cap (\cref{alg:split_comm}~\cref{alg:scomm}).
  In Step 2, the buffers prepared in Step 1 are sent to their destination node,
  specifically to the paired process on that node (\cref{alg:split_comm}~\cref{alg:gcomm}).
  For Step 3,
  all processes redistribute their received data to the
  correct destination processes on-node (\cref{alg:split_comm}~\cref{alg:rcomm}).}\label{fig:optstep}
\end{figure}

\Cref{alg:split_comm} provides the steps for performing Split communication once the relevant communicators
have been created. Depending on the computation in which Split communication is being
used,~\cref{alg:scomm,alg:gcomm,alg:rcomm}
of~\cref{alg:split_comm} can be overlapped with various pieces of the computation~---~details of overlapping
computation with node-aware communication can be found in~\cite{Bienz_napspmv}.
Furthermore, in~\cite{Lockhart_ecg_optna}, the
inter-node message size cutoff is determined by the rendezvous
protocol based on communication modeling for Lassen, but it is observed that the
message size cutoff can be determined via tuning or any other chosen criteria.
Similarly, we use a message size cutoff of three in~\cref{fig:optstep} to demonstrate
the multi-step technique when communicating between two nodes with four
processes each.

Splitting communication
eliminates the data redundancy from standard communication, but does so with
varying numbers of inter-node messages (as determined by the total data volume being
sent to another distinct node).
Within the context of a sparse matrix-block vector multiplication,
this scheme yields up to $60\times$ speedup over standard communication techniques.
The goal of this work is to consider approaches similar to the split communication strategy
within the context of heterogeneous architectures.

\subsection{Distributed Sparse Matrix-Vector Multiplication}\label{ss:spmv_background}
Throughout the paper, we utilize the irregular point-to-point communication patterns
induced by sparse matrix-vector multiplication (SpMV) to test the performance potential of
node-aware communication strategies within the context of GPU to GPU communication, as well
as provide further model validation.
A SpMV, defined as
\begin{equation}\label{eq:spmv}
   A \cdot v \rightarrow w
\end{equation}
with $A \in \Rmn$ and $v$, $w \in \Rn$, is a common kernel in sparse iterative methods.
Distributed SpMVs performed on GPUs currently face many performance hurdles
including computational inefficiencies of the local SpMV portion on each
GPU, packing and unpacking communication buffers,
strategically overlapping computation and communication,
etc.~\cite{jenkins2012enabling,osti_10064735}.
There are multiple potential solutions to these problems, many
of which are still currently being
researched~\cite{yang2018parallel,page2021scalability,buffer_packing}.

Because the presented work focuses on general communication strategies,
we do not attempt to optimize these portions of the distributed SpMV\@. Instead our
performance tests focus solely on benchmarking the irregular point-to-point communication
that occurs in the distributed kernel, characterizing the performance of
multiple communication strategies for various communicated data volumes and message counts
on a heterogeneous architecture.

\subsubsection{Testing Setup}
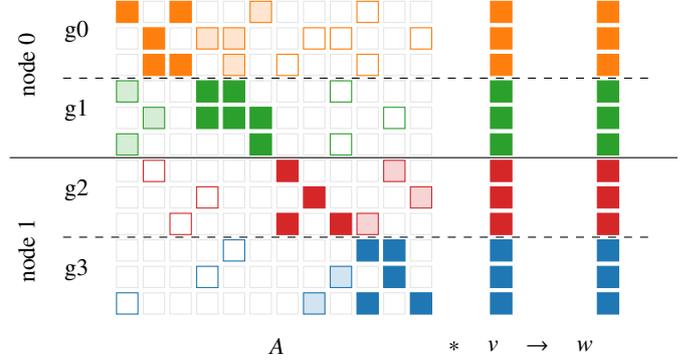
\begin{figure}[!ht]
    \centering
    \small %
    \definecolor{tab-blue}{HTML}{1f77b4}
\definecolor{tab-orange}{HTML}{ff7f0e}
\definecolor{tab-green}{HTML}{2ca02c}
\definecolor{tab-red}{HTML}{d62728}
\definecolor{tab-purple}{HTML}{9467bd}
\definecolor{tab-brown}{HTML}{8c564b}
\definecolor{tab-pink}{HTML}{e377c2}
\definecolor{tab-gray}{HTML}{7f7f7f}
\definecolor{tab-olive}{HTML}{bcbd22}
\definecolor{tab-cyan}{HTML}{17becf}

\newcommand{\aij}[5]
{
  \draw[fill=#4!#5,draw=#4] (#2/10+3/10, 1/10+#3/10-#1/10+0.01) rectangle +(0.08, 0.08);
}

\begin{tikzpicture}[x=100pt,y=100pt]
  \begin{scope}[shift={(10pt, 15pt)}]

  \foreach \z in {1, ..., 12}{%
    \foreach \y in {1,...,12}{%
      \def\x{\z/10+3/10}
      \draw[draw=tab-gray!20] (\x, \y/10+0.01) rectangle +(0.08, 0.08);
    }
  }

  \aij{1}{1}{12}{tab-orange}{100}
  \aij{2}{2}{12}{tab-orange}{100}
  \aij{3}{3}{12}{tab-orange}{100}
  \aij{1}{3}{12}{tab-orange}{100}
  \aij{3}{2}{12}{tab-orange}{100}

  \aij{4}{4}{12}{tab-green}{100}
  \aij{5}{5}{12}{tab-green}{100}
  \aij{6}{6}{12}{tab-green}{100}
  \aij{4}{5}{12}{tab-green}{100}
  \aij{5}{4}{12}{tab-green}{100}
  \aij{5}{6}{12}{tab-green}{100}

  \aij{7}{7}{12}{tab-red}{100}
  \aij{8}{8}{12}{tab-red}{100}
  \aij{9}{9}{12}{tab-red}{100}
  \aij{9}{7}{12}{tab-red}{100}

  \aij{10}{10}{12}{tab-blue}{100}
  \aij{11}{11}{12}{tab-blue}{100}
  \aij{12}{12}{12}{tab-blue}{100}
  \aij{10}{11}{12}{tab-blue}{100}
  \aij{12}{10}{12}{tab-blue}{100}

  \aij{1}{6}{12}{tab-orange}{20}
  \aij{2}{4}{12}{tab-orange}{20}
  \aij{2}{5}{12}{tab-orange}{20}
  \aij{3}{5}{12}{tab-orange}{20}

  \aij{4}{1}{12}{tab-green}{20}
  \aij{5}{2}{12}{tab-green}{20}
  \aij{6}{1}{12}{tab-green}{20}

  \aij{7}{11}{12}{tab-red}{20}
  \aij{8}{12}{12}{tab-red}{20}
  \aij{9}{10}{12}{tab-red}{20}

  \aij{11}{9}{12}{tab-blue}{20}
  \aij{12}{8}{12}{tab-blue}{20}

  \aij{1}{10}{12}{tab-orange}{0}
  \aij{2}{8}{12}{tab-orange}{0}
  \aij{2}{9}{12}{tab-orange}{0}
  \aij{2}{12}{12}{tab-orange}{0}
  \aij{3}{7}{12}{tab-orange}{0}
  \aij{3}{10}{12}{tab-orange}{0}

  \aij{4}{9}{12}{tab-green}{0}
  \aij{5}{11}{12}{tab-green}{0}
  \aij{6}{9}{12}{tab-green}{0}

  \aij{7}{2}{12}{tab-red}{0}
  \aij{8}{4}{12}{tab-red}{0}
  \aij{9}{3}{12}{tab-red}{0}

  \aij{10}{5}{12}{tab-blue}{0}
  \aij{11}{4}{12}{tab-blue}{0}
  \aij{12}{1}{12}{tab-blue}{0}

  \begin{scope}[shift={(1.4,0)}]
  \aij{1}{1}{12}{tab-orange}{100}
  \aij{2}{1}{12}{tab-orange}{100}
  \aij{3}{1}{12}{tab-orange}{100}
  \aij{4}{1}{12}{tab-green}{100}
  \aij{5}{1}{12}{tab-green}{100}
  \aij{6}{1}{12}{tab-green}{100}
  \aij{7}{1}{12}{tab-red}{100}
  \aij{8}{1}{12}{tab-red}{100}
  \aij{9}{1}{12}{tab-red}{100}
  \aij{10}{1}{12}{tab-blue}{100}
  \aij{11}{1}{12}{tab-blue}{100}
  \aij{12}{1}{12}{tab-blue}{100}
  \end{scope}

  \begin{scope}[shift={(1.8,0)}]
  \aij{1}{1}{12}{tab-orange}{100}
  \aij{2}{1}{12}{tab-orange}{100}
  \aij{3}{1}{12}{tab-orange}{100}
  \aij{4}{1}{12}{tab-green}{100}
  \aij{5}{1}{12}{tab-green}{100}
  \aij{6}{1}{12}{tab-green}{100}
  \aij{7}{1}{12}{tab-red}{100}
  \aij{8}{1}{12}{tab-red}{100}
  \aij{9}{1}{12}{tab-red}{100}
  \aij{10}{1}{12}{tab-blue}{100}
  \aij{11}{1}{12}{tab-blue}{100}
  \aij{12}{1}{12}{tab-blue}{100}
  \end{scope}

  \node[below, align=center] (a) at (1.0, 0.05) {$A$};
  \node[below, align=center, right= 0.55 of a] (b) {$*$};
  \node[below, align=center, right= .04 of b] (c) {$v$};
  \node[below, align=center, right= .04 of c] (d) {$\rightarrow$};
  \node[below, align=center, right= .04 of d] (e) {$w$};

  \draw[black, solid] (0, 0.70) -- (2.5, .70);
  \node[rotate=90,anchor=north] at (0.0, 1.05) {node $0$};
  \node[rotate=90,anchor=north] at (0.0, .35) {node $1$};

  \draw[black, dashed] (0.2, 1.0) -- (2.4, 1.0);
  \draw[black, dashed] (0.2, 0.4) -- (2.4, 0.4);
  \node[anchor=north] at (0.25, 1.25) {g0};
  \node[anchor=north] at (0.25, 0.95) {g1};
  \node[anchor=north] at (0.25, 0.65) {g2};
  \node[anchor=north] at (0.25, 0.35) {g3};

  \end{scope}
\end{tikzpicture}
    \caption{Partitioning of a SpMV, $A \cdot v\rightarrow w$, with $n=12$.
      Matrix $A$ and vectors $v$ and $w$ are partitioned across two nodes, four GPUs (g0, g1, g2, g3).
    Solid blocks,~\protect\tikz{\protect\draw[fill=black] (0,0) rectangle (0.2,0.2);},
    represent the portion of the SpMV requiring on-GPU values
    from $v$.
    Shaded blocks,~\protect\tikz{\protect\draw[fill=black!20] (0,0) rectangle (0.2,0.2);},
    require on-node but off-GPU communication of values from $v$.
    Outlined blocks,~\protect\tikz{\protect\draw[fill=black!0] (0,0) rectangle (0.2,0.2);},
    require values of $v$ from GPUs off-node.}\label{fig:spmv_nodeaware}
\end{figure}
All performance tests presented in~\cref{ss:model_validation,sec:spmv}
correspond to a distributed SpMV with $A$, $v$, and $w$ partitioned row-wise across
$g$ GPUs with contiguous rows stored on each GPU (see~\cref{fig:spmv_nodeaware}).
In addition, the rows of $A$ on each GPU are presumed to be split into two blocks,
namely on-GPU and off-GPU\@.
The on-GPU block is the diagonal block of columns corresponding to the
on-GPU portion of rows in $v$ and $w$, and the off-GPU block contains
the matrix $A$'s nonzero values correlating to non-local rows of $v$ and $w$ stored
off-GPU\@. This splitting is common practice, as it differentiates between the
portions of a SpMV that require communication, as well as making the distributed
kernel a perfect case study for node-aware communication performance on heterogeneous
architectures.
Because our key goal is to characterize irregular point-to-point communication performance
independent of the distributed operation in which it is included,
all presented benchmarks throughout the paper focus on the communication patterns
induced by the distributed SpMV and not the computational aspects of the operation.
We would like to note that within the context of a distributed SpMV, optimal performance
depends on some combination of communication and computation overlap. However, optimizing
the entire distributed SpMV operation lies outside the scope of this paper, thus
timings for the computational portion and on-device kernel details are excluded.

\section{Modeling Parameters for Communication}\label{sec:modeling_params}

When data is moved between two GPUs on separate nodes using MPI,
the data can be moved in one of two ways:
\begin{description}
    \item \textbf{Device-aware:} data is sent directly from the sending GPU
    through the NIC and the network to the receiving GPU without being copied
    to the host CPU\@; and
    \item \textbf{Staged-through-host:} data is copied to the host CPU before being
    sent through the NIC and the network to the receiving GPU's host CPU
    then copied to the receiving GPU\@.
\end{description}
Because both of these involve moving data through the GPU and possibly the CPU,
it is important to consider the cost of transmitting data through all possible
data flow paths involving the CPU or GPU\@.

Throughout this section and the remainder of the paper, results are presented
for the Spectrum MPI implementation on Lassen~\cite{lassen}
In~\cite{Bienz_modeling_hetero}, it is shown that Lassen and Summit~\cite{summit}
demonstrate similar performance using Spectrum MPI
(there, the MPI implementation is optimized for use on the two DOE machines),
therefore results for a single machine are provided.
Moreover, each of the presented model parameters is the result of ping-pong
and node-pong timings collected through
BenchPress\footnote{\url{https://github.com/bienz2/BenchPress}},
a node architecture-aware library used for benchmarking data movement
performance on large-scale systems. The ping-pong and node-pong tests are
performed for 1000 iterations and averaged;
each model parameter is then given by a linear least-squares fit to
the collected data.

We use the postal model presented in~\cref{eq:postal} to model the time required
for sending a single message between two CPUs or two GPUs, with the measured
parameters for Lassen presented in~\cref{table:postal}. The $\alpha$ and
$\beta$ parameters are separated based on where the two processes are physically
located with respect to one another, namely on the same socket, on different
sockets but the same node, or separate nodes. In addition, the parameters
are split further based on messaging protocol:
\begin{description}
\item[short] fits in the envelope
so the message is sent directly to the receiving process;
\item[eager] assumes
adequate buffer space is already allocated by the receiving process; or
\item[rendezvous] requires the receiving process to allocate buffer space for the
message before the data is sent.
\end{description}
The short protocol has been excluded from the GPU messaging parameter portion
of~\cref{table:postal} because this protocol is not used in device-aware
communication on Lassen.
\begin{table}[ht!]
    \renewcommand{\arraystretch}{1.1}
    \begin{center}
    \begin{tabular}{p{1mm} cr ccc}
      \toprule
        \multicolumn{3}{c}{} & on-socket & on-node & off-node \\
        \midrule
        \parbox[t]{1mm}{\multirow{6}{*}{\rotatebox[origin=c]{90}{inter-CPU}}}
        & \multirow{2}{*}{Short}
        & $\alpha$  & 3.67e-07 & 9.25e-07 & 1.89e-06 \\
        & & $\beta$ & 1.32e-10 & 1.19e-09 & 6.88e-10 \\
        \cline{2-6}
        & \multirow{2}{*}{Eager}
        & $\alpha$  & 4.61e-07 & 1.17e-06 & 2.44e-06 \\
        & & $\beta$ & 7.12e-11 & 2.18e-10 & 3.79e-10 \\
        \cline{2-6}
        & \multirow{2}{*}{Rend}
        & $\alpha$  & 3.15e-06 & 6.77e-06 & 7.76e-06 \\
        & & $\beta$ & 3.40e-11 & 1.49e-10 & 7.97e-11 \\
        \cline{1-6}
        \parbox[t]{1mm}{\multirow{4}{*}{\rotatebox[origin=c]{90}{inter-GPU}}}
        & \multirow{2}{*}{Eager}
        & $\alpha$  & 1.87e-06 & 2.02e-05 & 8.95e-06 \\
        & & $\beta$ & 5.79e-11 & 2.15e-10 & 1.72e-10 \\
        \cline{2-6}
        & \multirow{2}{*}{Rend}
        & $\alpha$  & 1.82e-05 & 1.93e-05 & 1.10e-05 \\
        & & $\beta$ & 1.46e-11 & 2.39e-11 & 1.72e-10 \\
     \bottomrule
    \end{tabular}
    \end{center}
    \hspace{3cm} $\alpha$ [sec] \qquad $\beta$ [sec/byte]
    \caption{Measured parameters for inter-CPU and inter-GPU communication on Lassen.}\label{table:postal}
\end{table}

\begin{figure}
    \centering
    \includegraphics[width=\columnwidth]{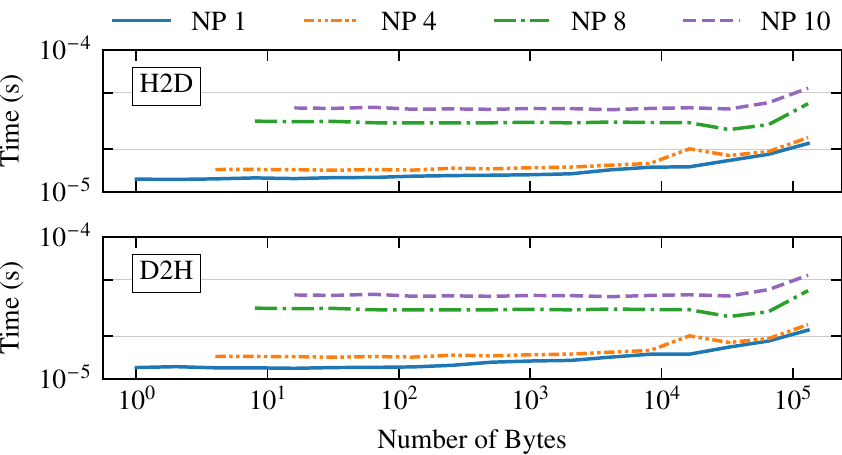}
    \caption{The time required to copy various amounts of data from a single
    GPU using \texttt{cudaMemcpyAsync} when splitting the copy across
    NP processes. HostToDevice (H2D) and DeviceToHost (D2H) timings presented.}\label{fig:memcpy_mult_split}
\end{figure}

Because staging data through a host process requires copying data to the sending
host CPU and from the receiving GPU's host process,
measured parameters for \texttt{cudaMemcpyAsync} are included
in~\cref{table:memcpy} with distinction between whether the copy is using
a single process or four processes to move data from the device.
We assume that all data copies will occur on-socket, and we do not consider
cases with more than four processes pulling data from a single GPU at a time
as there was no observed benefit in splitting data copies further across multiple
processes~---~see~\cref{fig:memcpy_mult_split}.
\begin{table}[ht!]
    \renewcommand{\arraystretch}{1.1}
    \begin{center}
    \begin{tabular}{rc cc}
      \toprule
      \multicolumn{2}{c}{} & HostToDevice & DeviceToHost \\
      \midrule
        \multirow{2}{*}{1 proc}
        & $\alpha$ & 1.30e-05 & 1.27e-05\\
        & $\beta$  & 1.85e-11 & 1.96e-11 \\
        \cline{1-4}
        \multirow{2}{*}{4 proc}
        & $\alpha$ & 1.52e-05 & 1.47e-05\\
        & $\beta$  & 5.52e-10 & 1.50e-10\\
        \bottomrule
    \end{tabular}
    \end{center}
    \hspace{3cm} $\alpha$ [sec] \qquad $\beta$ [sec/byte]
    \caption{Measured parameters for \texttt{cudaMemcpyAsync} on Lassen.}\label{table:memcpy}
\end{table}

In addition to considering the postal model for inter-CPU and inter-GPU
communication, the max-rate model presented in~\cref{eq:max_rate} is required
for accurately predicting the performance of staging GPU data through a
host process when using more than a single process per node. Therefore,
the measured injection bandwidth limit for inter-CPU communication is
presented in~\cref{table:maxrate}.
The inter-GPU injection bandwidth limit is excluded, as these limits are not
reached with the four available GPUs per node on Lassen.
\begin{table}[ht!]
    \centering
    \begin{tabular}{r c}
      \toprule
     \multicolumn{1}{c}{} & $R_{N}^{-1}$ [bytes/sec]\\
     \midrule
    inter-CPU &  4.19e-11\\
    \bottomrule
    \end{tabular}
    \medskip
    \caption{Measured parameter for injection bandwidth limits on Lassen.}\label{table:maxrate}
\end{table}
Using the measured modeling parameters, we now model the performance of
various communication strategies based on the node-aware techniques discussed
in~\cref{ss:na}.

\section{Modeling Node-Aware Strategies for Inter-Node Communication}\label{sec:na-hetero}

In this section, we present performance models for existing node-aware
strategies using device-aware and staged-through-host communication for inter-node
data exchanges on Lassen, though these models do extend to any machine with
two sockets per node.
For each node-aware case, the models are divided into the time spent in
on-node communication (\cref{ss:on_node23,ss:on_node_opt}),
off-node communication (\cref{ss:model_inter_node}),
and data copies, in the case of staged-through-host communication (see~\cref{ss:copy}).
The models themselves do not consider the
removal of duplicate data discussed in~\cref{ss:na}, as the amount
of duplicate data injected into the network is operation and
problem dependent. However, adapting the input parameters for the models to
reflect the removal of duplicate data is straightforward and done
in~\cref{ss:model_plots}.

Performance is modeled for standard communication and all node-aware communication
strategies discussed in~\cref{ss:na}. We consider staged-through-host and device-aware
communication for all of the strategies except for the split strategies, for
which device-aware communication does not apply.
``Split + MD'' first copies data to a single host process, then splits the inter-node data
to be communicated across multiple processes via extra on-node inter-CPU messages.
``Split + DD'' uses duplicate device pointers to copy data from a GPU to
multiple host process, reducing the number of on-node messages required to split
the inter-node data volume being communicated. Each GPU is assumed to have
a single host process except in the case of ``Split + DD''\@.
For reference, the modeled communication strategies are listed in~\cref{table:comm_strategies}.
\begin{table}[ht!]
    \renewcommand{\arraystretch}{1.2}
    \begin{center}
    \begin{tabular}{c cc}
      \toprule
        \multicolumn{1}{c}{} & \textbf{Staged-through-host} & \textbf{Device-aware} \\
      \midrule
        Standard  & \checkmark\ & \checkmark\ \\
        \cline{1-3}
        3-Step    & \checkmark\ & \checkmark\ \\
        \cline{1-3}
        2-Step    & \checkmark\ & \checkmark\ \\
        \cline{1-3}
        Split + MD & \checkmark\ & \\
        \cline{1-3}
        Split + DD & \checkmark\ &  \\
        \bottomrule
    \end{tabular}
    \end{center}
    \caption{Modeled communication strategies.}\label{table:comm_strategies}
\end{table}

\subsection{Modeling On-Node Communication for 3-Step and 2-Step}\label{ss:on_node23}

For 3-Step communication,
all data originating on any GPU on node $k$ with a destination of any
GPU on node $l$ is first gathered locally.
In the worst-case scenario,
all GPUs on node $k$ must contribute data for node $l$, requiring
communication among all GPUs per node.
This is modeled as
\begin{equation}\label{eq:on}
\begin{split}
  T_{\texttt{on}}(s) & = (\texttt{gps} -1) (\alpha_{\texttt{on-socket}} +
                    \beta_{\texttt{on-socket}}\cdot s)  \\
                       & + \texttt{gps} \cdot (\alpha_{\texttt{on-node}} +
                    \beta_{\texttt{on-node}}\cdot s)
\end{split}
\end{equation}
where \texttt{gps} is the GPUs per socket and
$s$ is the maximum message size sent by any single GPU\@.

The last step of both 2-Step and 3-Step communication involves redistributing
data received via inter-node communication to its final destination GPU on-node.
The worst case scenario for both strategies occurs when all of the data received
via inter-node communication needs to be redistributed to every other
GPU on-node. This is also modeled with~\cref{eq:on}, with $s$ representing the
maximum received inter-node message size.

\subsection{Modeling On-Node Communication for Split}\label{ss:on_node_opt}
The split strategies require copying all
data on node $k$ with destination of any GPU on node $l \neq k$ to the
host processes before distributing the data across some number
of on-node processes. Finally, each process sends data through the network.
For large inter-node message sizes, the worst-case scenario occurs when a single GPU
contains all data to be sent off-node with a data size large enough that it is
split across all on-node processes.
In the case of Lassen, there are a maximum of 40 on-node processes,
therefore distributing the data would require an additional 19 on-socket messages
and 20 off-socket/on-node messages if a single host process per GPU were being used.
Generalizing the split strategy to any architecture using multiple host processes with duplicate
device pointers yields
\begin{equation}\label{eq:on_opt}
\begin{split}
  T_{\texttt{on-split}}(s,\texttt{ppg}) & =  \left(\frac{\texttt{pps}}{\texttt{ppg}}-1\right) \cdot
            (\alpha_{\texttt{on-socket}} + \beta_{\texttt{on-socket}} \cdot s) \\
                    & + \left(\frac{\texttt{pps}}{\texttt{ppg}}\right) \cdot
    (\alpha_{\texttt{on-node}} + \beta_{\texttt{on-node}} \cdot s)
\end{split}
\end{equation}
as the modeled time, where
\texttt{ppg} is the number of host processes per GPU,
and \texttt{pps} is the processes per socket, and $s$ is the total data
volume to be communicated inter-node split across \texttt{ppg}.

Similar to the worst-case scenario for 3-Step and 2-Step on-node communication,
the worst-case redistribution scenario for the split strategies is equivalent
to~\cref{eq:on_opt}.  In this case,
a single GPU must redistribute all received inter-node
data to every other GPU on-node; here,
$s$ represents
the total data volume received via inter-node communication split across
\texttt{ppg}.

\subsection{Modeling Off-Node Communication}\label{ss:model_inter_node}

For the off-node communication portion of each of the multi-step communication
strategies, the max-rate model~\cref{eq:max_rate} is used for
routines that are staged-through-host, and the postal model~\cref{eq:postal} is used for device-aware routines.
For the max-rate model, the time spent in off-node communication is given by
\begin{equation}\label{eq:off}
  T_{\texttt{off}}(m,s) = \alpha_{\texttt{off-node}} \cdot m +
            \text{max} \left(\frac{s_\texttt{node}}{R_N}, s \cdot \beta_{\texttt{off-node}}\right)
\end{equation}
for a number of messages to be communicated, $m$, and a maximum number of
bytes sent by a single process, $s_{\texttt{proc}}$ where
$s_{\texttt{node}}$ is the maximum number of bytes injected into the network by any single
node.
For device-aware communication, this reduces to the postal model
\begin{equation}\label{eq:off_da}
  T_{\texttt{off-DA}}(m,s) = \alpha_{\texttt{off-node}} \cdot m
                    + s \cdot \beta_{\texttt{off-node}}.
\end{equation}
\begin{table*}[t!]
    \renewcommand{\arraystretch}{1.25}
    \begin{center}
    \begin{tabular}{cc c}
      \toprule
        \multicolumn{2}{c}{Communication Strategy} & Model \\
      \midrule
        \multirow{2}{*}{Standard} & Staged-through-host & Max-rate model~\cref{eq:max_rate} \\
          & Device-aware & Postal model~\cref{eq:postal} \\
        \cline{1-3}
        \multirow{2}{*}{3-Step} & Staged-through-host &
            $ T_\texttt{off}(m_{\texttt{node}\rightarrow\texttt{node}},
                s_{\texttt{node}\rightarrow\texttt{node}}) +
            2 \cdot T_\texttt{on}(s_{\texttt{node}\rightarrow\texttt{node}}) +
            T_{\text{copy}}(s_\texttt{proc},s_{\texttt{node}\rightarrow\texttt{node}})$ \\
          & Device-aware &
          $ T_\texttt{off-DA}(m_{\texttt{node}\rightarrow\texttt{node}},
                s_{\texttt{node}\rightarrow\texttt{node}}) +
          2 \cdot T_\texttt{on}(s_{\texttt{node}\rightarrow\texttt{node}})$ \\
        \cline{1-3}
        \multirow{2}{*}{2-Step} & Staged-through-host &
            $ T_\texttt{off}(m_{\texttt{proc}\rightarrow\texttt{node}},s_{\texttt{proc}}) +
            T_\texttt{on}(s_{\texttt{proc}}) +
            T_{\texttt{copy}}(s_\texttt{proc},s_{\texttt{node}\rightarrow\texttt{node}})$ \\
          & Device-aware &
            $ T_\texttt{off-DA}(m_{\texttt{proc}\rightarrow\texttt{node}},s_{\texttt{proc}}) +
            T_\texttt{on}(s_{\texttt{proc}})$ \\
        \cline{1-3}
        \multirow{2}{*}{Split} & Staged-through-host + MD &
            $ T_\texttt{off}(m_{\texttt{proc}\rightarrow\texttt{node}},
            s_{\texttt{node}} / _\texttt{ppn}) +
            2 \cdot T_\texttt{on-split}(s_{\texttt{node}},1)
            + T_{\texttt{copy}}(s_\texttt{proc},s_{\texttt{node}\rightarrow\texttt{node}})$ \\
          & Staged-through-host + DD &
            $ T_\texttt{off}(m_{\texttt{proc}\rightarrow\texttt{node}},
            s_{\texttt{node}} / _\texttt{ppn}) +
            2 \cdot T_\texttt{on-split}(s_{\texttt{node}},4)
            + T_{\texttt{copy}}(s_\texttt{proc},s_{\texttt{node}\rightarrow\texttt{node}})$ \\
        \bottomrule
    \end{tabular}
    \end{center}
    \caption{Communication models. (Extra parameters defined in~\cref{table:extra_params})}\label{table:comm_models}
\end{table*}

\subsection{Copy Parameter for Staged-through-Host Communication}\label{ss:copy}

The time required to copy data between the CPU
and GPU is given by
\begin{equation}\label{eq:copy}
\begin{split}
  T_{\texttt{copy}}(s_\texttt{send}, s_\texttt{recv}) = &\alpha_{\texttt{H2D}} +
                    \beta_{\texttt{H2D}}\cdot s_{\texttt{send}} \\
                    & + \alpha_{\texttt{D2H}} +
                    \beta_{\texttt{D2H}}\cdot s_{\texttt{recv}}.
\end{split}
\end{equation}
where $s_{\texttt{send}}$ is the initial data copied from the source GPU,
and $s_{\texttt{recv}}$ is the final data copied to the destination GPU\@.

For all communication strategies except splitting with duplicate device pointers,
a single process copies all data from a corresponding GPU\@. In the case of splitting
with duplicate device pointers, we set the number of processes copying data
simultaneously to four in our model. Parameters for both a single host process copying
data and four host processes copying data simultaneously are presented
in~\cref{table:memcpy}.

\subsection{Model Validation}\label{ss:model_validation}
\begin{table}[ht!]
    \renewcommand{\arraystretch}{1.1}
    \begin{center}
    \begin{tabular}{r l}
      \toprule
        \multicolumn{1}{c}{Parameter} & Description \\
      \midrule
        $s_{\texttt{proc}}$ & {\small max \# of bytes sent by a single process/ GPU} \\
        $s_{\texttt{node}}$ & {\small max \# of bytes injected by a single node} \\
        $s_{\texttt{node}\rightarrow\texttt{node}}$ &
            {\small max \# of bytes sent between any two nodes} \\
        $m_{\texttt{proc}\rightarrow\texttt{node}}$ &
            {\small max \# of nodes to which a processor sends} \\
        $m_{\texttt{node}\rightarrow\texttt{node}}$ &
            {\small max \# of messages between any two nodes} \\
        \bottomrule
    \end{tabular}
    \end{center}
    \caption{Extra modeling parameters.}\label{table:extra_params}
\end{table}
\Cref{table:comm_models} presents the full models for the various communication
strategies given in~\cref{table:comm_strategies}, which combine the preceding
sub-models, with extra model parameters defined in~\cref{table:extra_params} for clarity.
\begin{figure}[ht!]
    \centering
    \includegraphics[width=0.5\columnwidth]{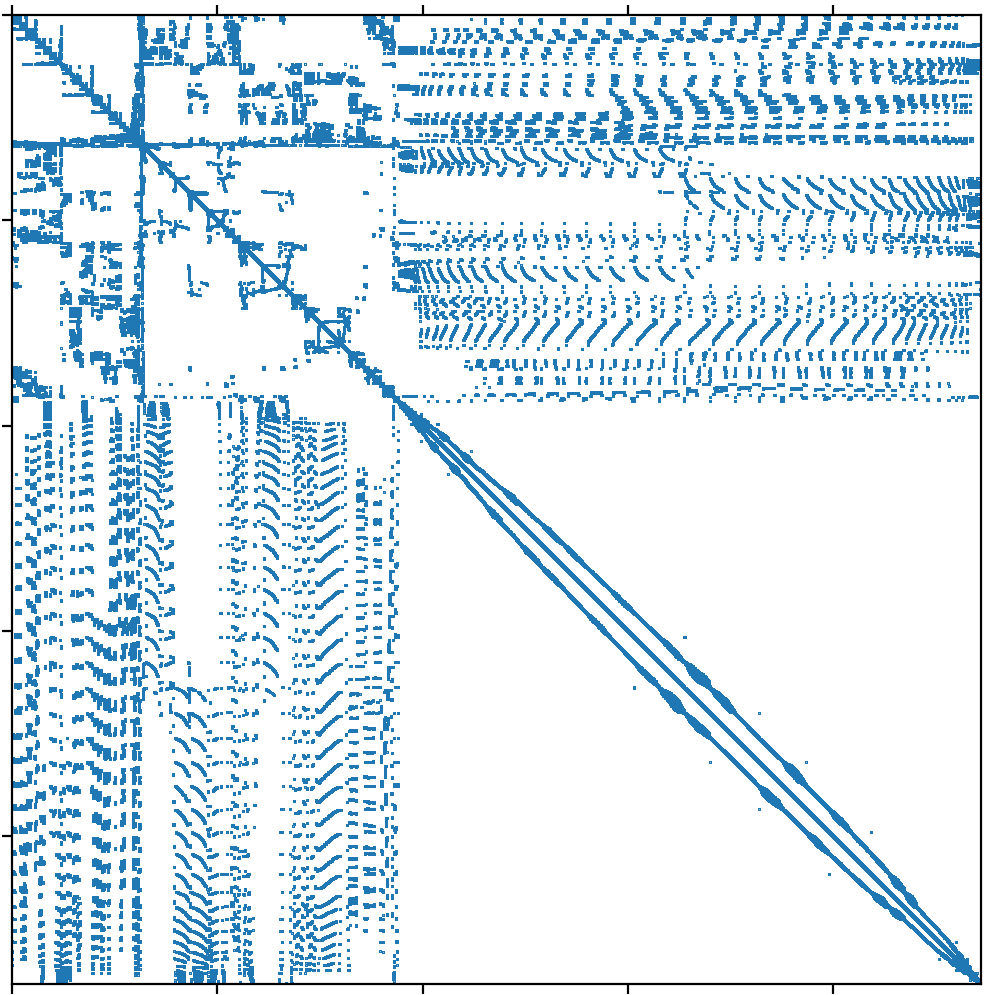}
    \caption{Sparsity pattern for the \texttt{audikw\_1} matrix.}\label{fig:sparsity}
\end{figure}

We provide a brief validation of the models via performance of the communication pattern induced by
sparse matrix-vector multiplication (SpMV) with the \texttt{audikw\_1} matrix from the Suite\-Sparse
matrix collection~\cite{ufl_matrices}. The matrix has \num{943695} rows and columns, and a nonzero
density of 8.72e-05 with the associated sparsity pattern in~\cref{fig:sparsity}.
Due to the high number of nonzero entries in the top rows and first columns of the matrix,
the communication pattern associated with a SpMV for \texttt{audikw\_1}
incurs high numbers of on-node and inter-node communication, therefore it is a perfect
test case for validating the models which model the worst-case on-node communication
scenarios for each of the communication strategies.

\begin{figure}[ht!]
    \centering
    \includegraphics[width=\columnwidth]{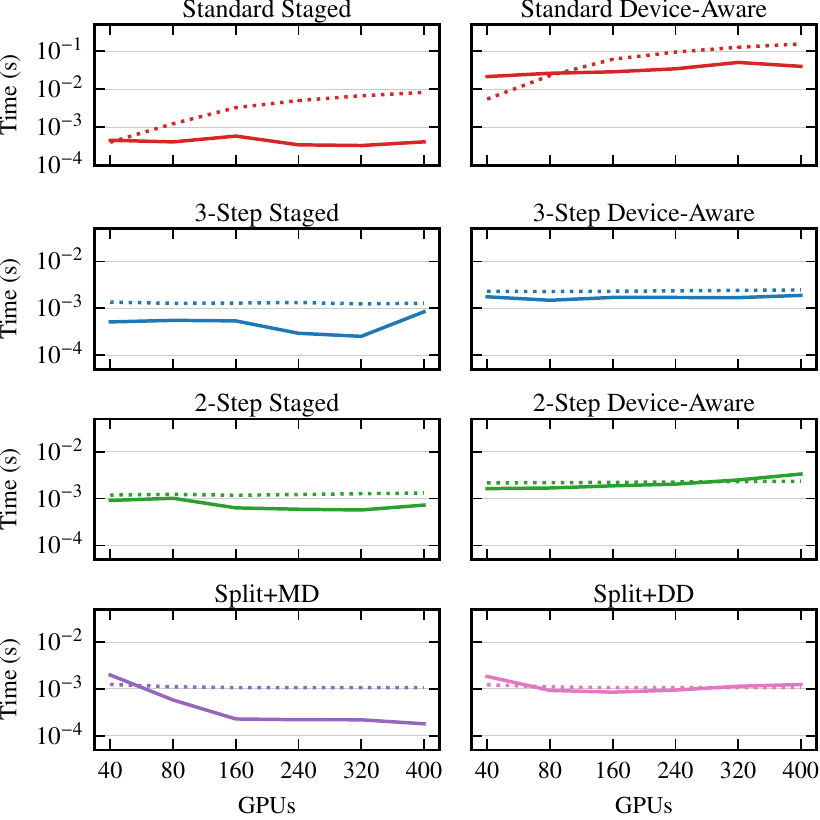}
    \caption{Model validation.
    Solid lines,~\protect\tikz{\protect\draw[solid,thick] (0,0)--(0.5,0);}, %
    depict measured times, and dotted lines,~\protect\tikz{\protect\draw[dotted,thick] (0,0)--(0.5,0);}, %
    depict model predictions.}\label{fig:validation}
\end{figure}

\Cref{fig:validation} depicts the measured times (solid lines) for SpMV communication alongside model
predictions (dotted lines). Presented measured times are the \textit{maximum} average time required
for communication by any \textit{single} process for 1000 test runs.
In the standard communication cases, the modeled times are an order of
magnitude higher than actual measured times, but for the node-aware commmunication
models, the predicted times provide a tight upperbound, generally on the same order of
magnitude as the measured performance.
In~\cref{ss:model_plots}, we use these models to predict the performance of
common irregular point-to-point communication scenarios.

\subsection{Modeled Performance}\label{ss:model_plots}
\begin{figure*}[p!]
    \centering
    \begin{subfigure}[b]{\textwidth}
        \centering
        \includegraphics[width=\textwidth]{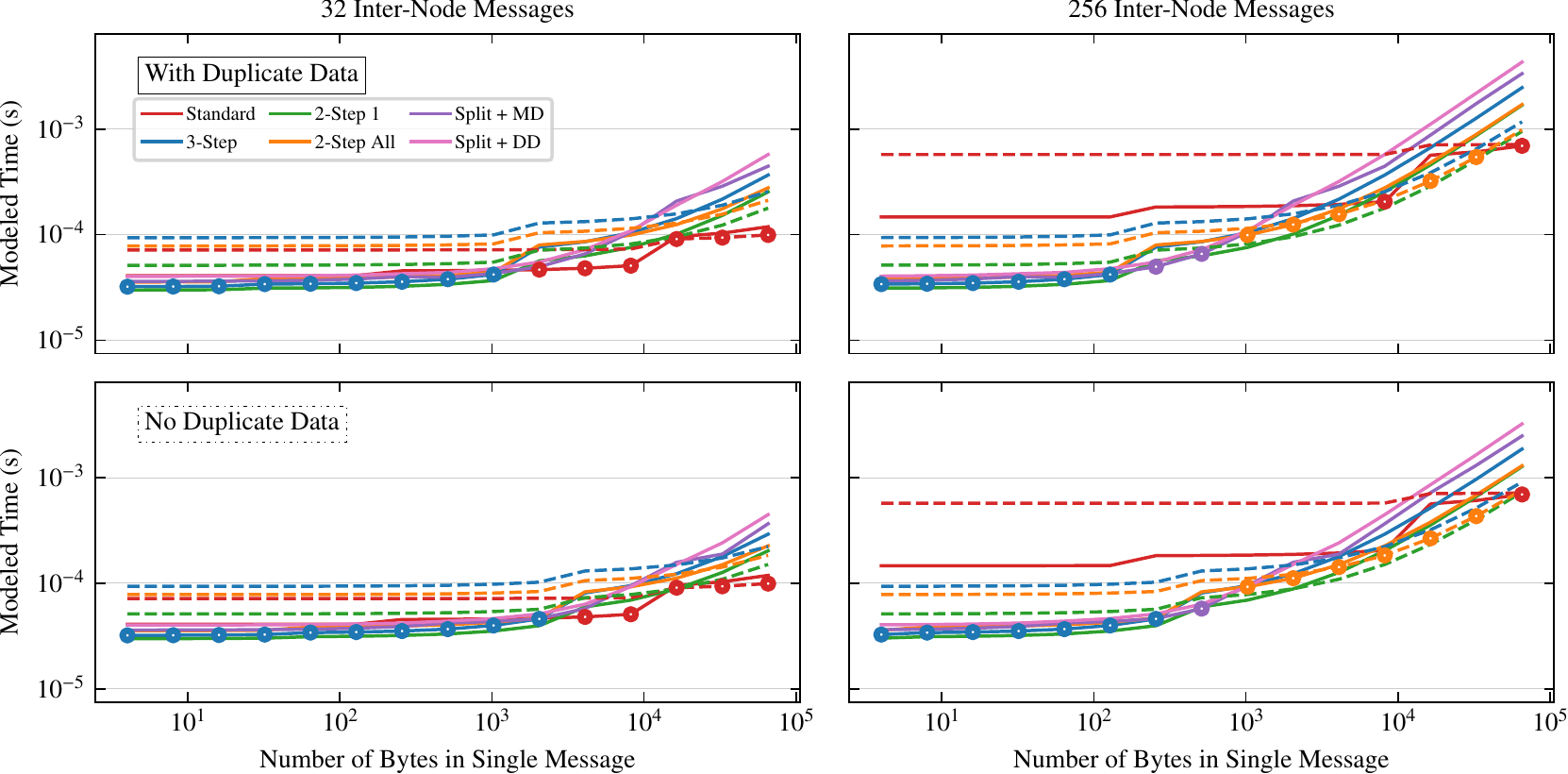}
        \caption{4 Nodes}\label{subfig:models_4nodes}
    \end{subfigure}
    \begin{subfigure}[b]{\textwidth}
        \vspace{1em}
        \centering
        \includegraphics[width=\textwidth]{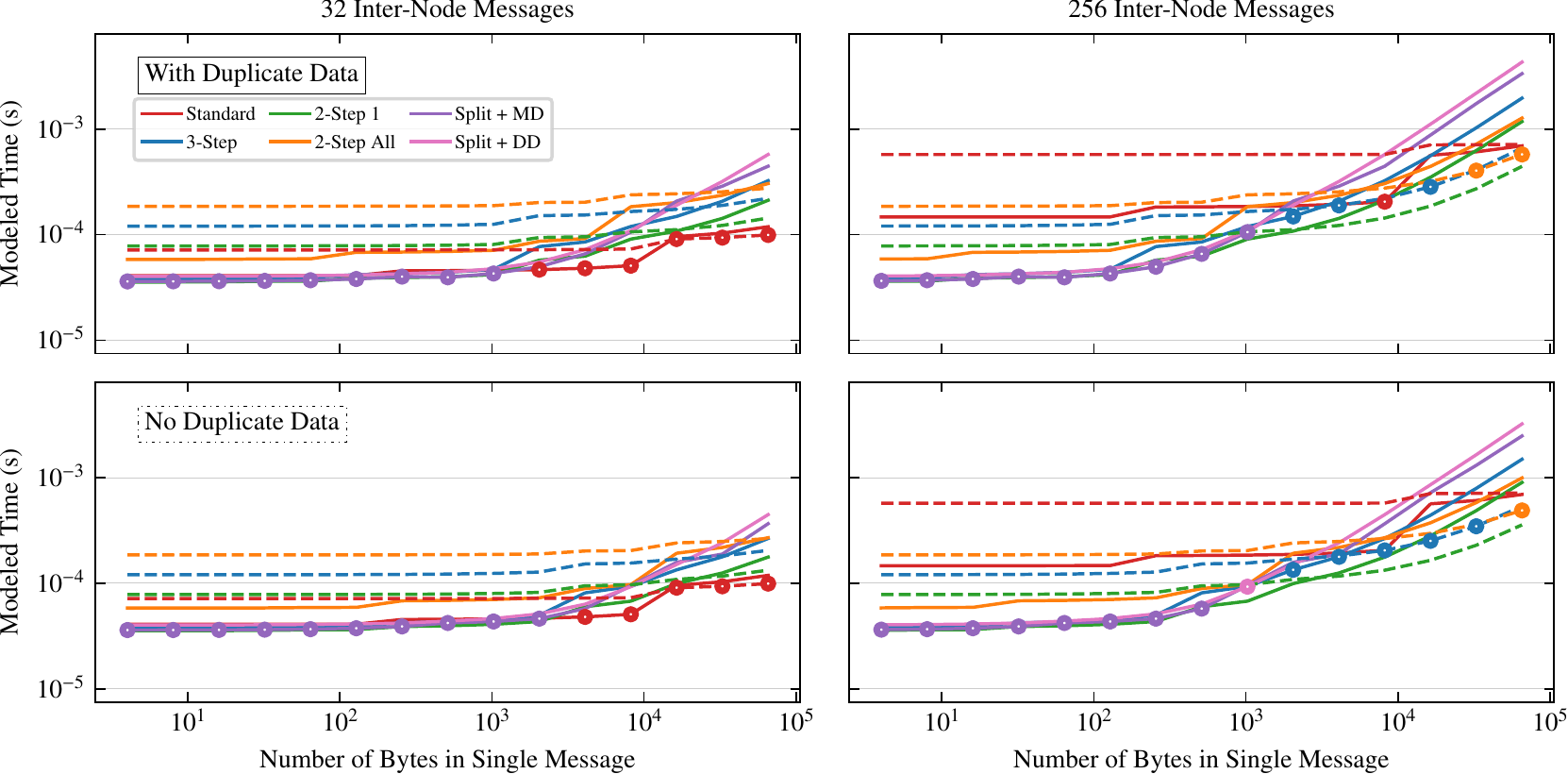}
        \caption{16 Nodes}\label{subfig:models_16nodes}
    \end{subfigure}
    \caption{The modeled time to send data from a single node to \textbf{4 nodes (top)} and
    \textbf{16 nodes (bottom)}, where data from the sending node is sent via 32 or
    256 messages distributed evenly across all on-node GPUs when using Standard
    communication. The bottom rows depict the scenario when 25\% of the original data was `duplicate data' and is removed.
    Solid lines,~\protect\tikz{\protect\draw[solid,thick] (0,0)--(0.5,0);}, %
    depict staged-through-host communication, and dashed lines,~\protect\tikz{\protect\draw[dashed,thick] (0,0)--(0.5,0);}, %
    depict device-aware communication. The minimum modeled times (excluding the 2-Step \textit{best-case} scenario, 2-Step 1) are bolded.}\label{fig:models}
\end{figure*}

\Cref{fig:models} presents the modeled performance for common scenarios with
irregular point-to-point communication, namely, a node sending a modest number
of inter-node messages (32) and a large number of inter-node messages (256),
with messages distributed evenly across on-node GPUs.
Because the node-aware performance models are dependent upon the number
of destination nodes, the models are split further, modelling if the
data was being sent to 4 nodes (\cref{subfig:models_4nodes}) or
16 nodes (\cref{subfig:models_16nodes}). Additionally, the bottom
rows of both~\cref{subfig:models_4nodes} and~\cref{subfig:models_16nodes}
depict the impact of duplicate data removal on the node-aware
communication strategies, by modelling their performance if 25\% of
the original data was duplicate data.

For each of these scenarios,
we model the amount of time required for each node to send their
messages to the destination nodes using standard communication. This
modeled performance is compared against that of the various node-aware
strategies where the messages are split and/or agglomerated accordingly.
There are two cases presented for the 2-Step strategy,
considering if every GPU on the source node is sending data to every GPU
on the destination node (2-Step All),
or if all the messages being sent to the destination node are from
a single active GPU on the source node (2-Step 1).
The message size cap for the split strategies is taken to be the same
that was used in~\cite{Lockhart_ecg_optna} and is the
message cap used for switching to the rendezvous protocol on Lassen.

In~\cref{fig:models}, we have circled the minimum modeled times for convenience, excluding the 2-Step 1 approaches, as they present the \textit{best-case}
scenario for 2-Step communication, which does not often occur in practice. However, we do think it is important to present these modeled times
in order to depict a comprehensive picture of node-aware communication's potential.
For large message counts (256 Inter-Node Messages plots in~\cref{fig:models}) and for message sizes greater than $10^3$ Bytes,
device-aware 2-Step 1 is predicted to perform best, indicating that for high inter-node data volumes, if the on-node data was distributed such that every
GPU on a given node $k$ was communicating with a distinct GPU on destination node $l$, 2-Step communication would be best.
This is consistent with the observed performance of the application-specific hierarchical communication in~\cite{petascale_xct}.
Now, we include discussion of the circled minimum times excluding the 2-Step 1 performance predictions.

In the case of
a small number of messages injected into the network to a small number of nodes (\cref{subfig:models_4nodes}),
3-Step and standard communication are observed as the most performant with Split+MD communication replacing 3-Step as the most
performant for 16 nodes (\cref{subfig:models_16nodes}). In both cases, the staged-through-host strategies predict
the best performance until message sizes grow extremely large ($> 10^4$ Bytes), where standard device-aware communication
is modeled to be best. Device-aware communication is also modeled to be best for large message sizes when a node is injecting
a large number of messages into the network. However, due to the high message volume, 3-Step and 2-Step device-aware
strategies are predicted to have the optimal performance, due to their reduction in messages sent compared to
standard communication.

Staged-through-host node-aware communication techniques model the best performance independent of number of destination nodes for all
message sizes up to $10^4$ Bytes.
When communicating with a small number of nodes (\cref{subfig:models_4nodes}), 3-Step and 2-Step communication are often
predicted to be the most performant, while Split+MD communication is predicted to be the most performant when communicating with a
larger number of nodes (\cref{subfig:models_16nodes}). This can be attributed to the use of all available processes on-node (40
in the case of Lassen), so that each individual process is injecting fewer messages into the network than in the case of 3-Step or 2-Step communication,
where there is only a single process paired with each GPU (4 in the case of Lassen).

The device-aware node-aware strategies models present relatively large costs.
However, this is unsurprising, considering the high overhead for
inter-GPU communication on-socket and on-node (as indicated by the
measured parameters in~\cref{table:postal}). The only cases
for which device-aware node-aware strategies have improved performance over staged-through-host techniques
is when the communicated inter-node data volume is \textit{extremely} large, or assumed to have an
optimal communication pattern (as in 2-Step 1).

Concerning the removal of duplicate data, there should be no impact on performance for small numbers of inter-node messages.
A performance impact is noticed primarily when their is communication of a high inter-node data volume via a high number of
messages. Once message sizes grow past $10^3$ Bytes in standard communication for all high message count models, removing
duplicate data impacts which node-aware communication strategy could be most performant, switching from Split + MD to Split + DD
or 2-Step to 3-Step, in the case of~\cref{subfig:models_16nodes}.

Overall, the staged-through-host node-aware communication strategies model the
best predicted performance for communication patterns requiring
a high number of inter-node message exchanges. In~\cref{sec:spmv},
we benchmark the performance of the communication strategies within the context
of sparse matrix-vector multiplication, verifying model predictions.

\section{Benchmarking Sparse Matrix-Vector Multiplication Communication Patterns}\label{sec:spmv}
In this section, we present performance results for the various
communication strategies discussed throughout~\cref{sec:modeling_params,sec:na-hetero}
when applied to the communication patterns of a single distributed SpMV~--~see~\cref{ss:spmv_background}\@.
For each of the strategies, each GPU is assumed to have a single host process, except
in the case of ``Split + DD'' where four host processes are used. Additionally, for the
split strategies, inter-node communicated data is potentially partitioned across up to
40 processes on-node (the maximum number of on-node processes for Lassen.)
Our test matrices are a subset of the largest matrices in the Suite\-Sparse matrix
collection~\cite{ufl_matrices}.
For each benchmark, we performed 1000 test runs and present the \textit{maximum}
average time required for communication by any \textit{single} process.

\subsection{Results}
\begin{figure*}[p!]
    \centering
    \includegraphics[width=\textwidth]{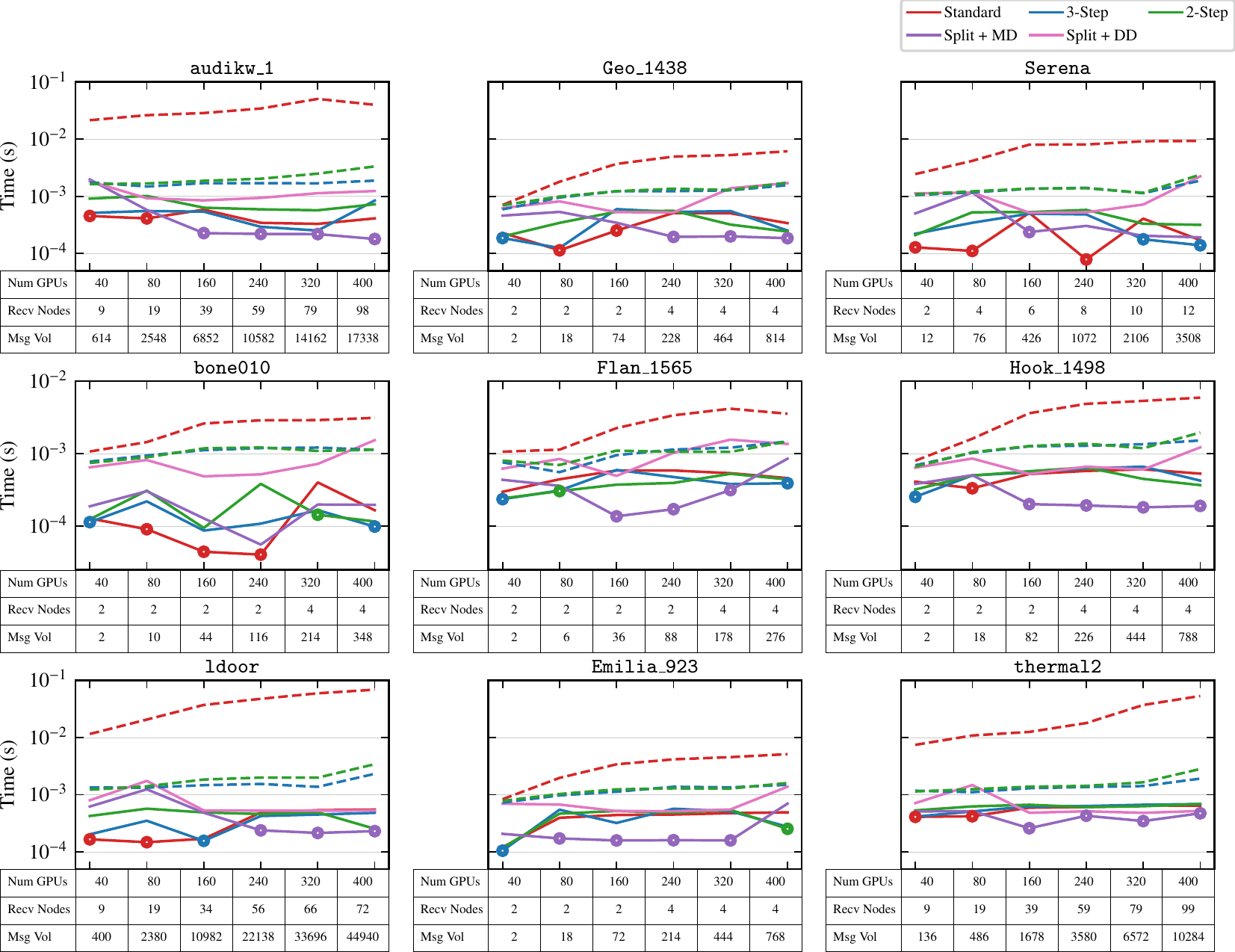}
    \caption{The measured time spent in irregular point-to-point communication for a
    distributed SpMV for various Suite\-Sparse matrices. Number of GPUs across which
    the problem was partitioned and standard communication maximum number of connected nodes
    for any single node (Recv Nodes) and message volume included beneath plots.
    Solid lines,~\protect\tikz{\protect\draw[solid,thick] (0,0)--(0.5,0);}, %
    depict staged-through-host communication, and dashed lines,~\protect\tikz{\protect\draw[dashed,thick] (0,0)--(0.5,0);}, %
    depict device-aware communication.
    Minimum times circled for convenience.}\label{fig:na_benchmarks}
\end{figure*}
~\cref{fig:na_benchmarks} displays the distributed SpMV communication benchmark
performance times for each communication strategy presented in~\cref{sec:na-hetero}
for each Suite\-Sparse matrix.
Presented beneath each plot is the number of GPUs on which the SpMV is partitioned,
the maximum number of nodes to which any single node is communicating (Recv Nodes),
and the communicated inter-node message volume for standard communication.

Consistent with the majority of model predictions for large inter-node message
volumes, the staged-through-host communication strategies exhibit far faster performance
than the device-aware communication strategies. However, it is worth noting that
device-aware 3-Step and device-aware 2-Step are typically much faster than standard
device-aware communication. In the case of the \texttt{thermal2} matrix, which exhibited a
high inter-node message volume for standard communication, the
gap between the device-aware node-aware strategies and staged-through-host communication
strategies is smaller than for other matrices. Additionally, ``Split + DD''
consistently performed worse than ``Split + MD'', consistent with modeled predictions.
This is unsurprising considering the latency associated with using duplicate device
pointers ($\mathtt{\sim}$1.5e-05 in~\cref{table:memcpy}) is much higher than the latency of
sending on-socket messages
($\mathtt{\sim}$3.7e-07--$\mathtt{\sim}$3.2e-06 in~\cref{table:postal}) to
distribute data being sent from a single GPU across multiple ranks.

The majority of the presented results are similar to the model prediction plots (\cref{fig:models}),
where the fastest communication
strategy was typically predicted to be one of the staged-through-host strategies: ``Split + MD'', Standard, or ``3-Step''.
``Split + MD'' exhibits the minimal performing time in most cases, except for smaller counts of participating
GPUs (40 or 80 in the case of \texttt{audikw\_1}, \texttt{Serena}, \texttt{ldoor}, \texttt{thermal2}), or for low
inter-node message counts (\texttt{bone010}, \texttt{Geo\_1438}) in which standard communication becomes
more optimal.

Overall, staged-through-host node-aware communication strategies demonstrate the best
performance for the majority of benchmarks, with ``Split + MD'' typically being the
fastest, consistent with model predictions in~\cref{subfig:models_16nodes}.

\section{Conclusions and Future Work}\label{sec:conclusion}
The advancement of parallel computers has introduced the design of supercomputers
with heterogeneous compute nodes due to the inclusion of multiple GPUs per node.
For distributed applications, this typically results in larger communicated data volumes,
as each compute unit can now operate on a larger partition of the problem.
In addition to increased data volumes, the inclusion of multiple GPUs per node
has increased the complexity of determining optimal data movement paths, particularly
in the case of inter-node irregular point-to-point communication.
In this work, we characterized the performance of irregular point-to-point communication
between GPUs via modeling and introduced node-aware communication strategies to
inter-node communication on heterogeneous architectures.
Our models suggested the use of staged-through-host
node-aware communication strategies, specifically split methods were indicated
as potential top performers. These results were confirmed by a performance study
on distributed SpMVs which saw split node-aware communication performing best
in most cases, and typically much faster than standard device-aware communication.

Additionally, our work provides important groundwork on designing efficient communication
strategies for the next generation of supercomputers.
Future exascale machine architectures will include higher CPU core counts per node,
alongside higher bandwidth interconnects (e.g., on Frontier, El Capitan, or Delta), two parameters that largely
affect the performance of node-aware communication strategies. Based on the
presented models, Split communication strategies will likely be the most efficient communication techniques to take advantage
of the high bandwidth interconnects, but distributing data to be communicated across
a larger number of on-node CPU cores could pose performance constraints.
Because the models presented in~\cref{sec:na-hetero} naturally extend to
architectures with single socket nodes, future work includes plans to begin modeling the
performance of machines resembling the next generation DOE exascale machines.

\section*{Acknowledgements}
This material is based in part upon work supported by the Department of
Energy, National Nuclear Security Administration, under Award Number
\textit{DE-NA0003963} and \textit{DE-NA0003966}.

\clearpage

\bibliographystyle{elsarticle-num}
\bibliography{optcomm.bib}

\end{document}